\documentclass[a4paper,11pt]{article}
\usepackage{jheppub}
\usepackage{graphicx} 
\usepackage{comment}
\usepackage{xcolor}
\usepackage{hyperref}
\hypersetup{
    colorlinks=true,
    linkcolor=blue,
    filecolor=cyan,      
    urlcolor=purple,
    pdftitle={Overleaf Example},
    pdfpagemode=FullScreen,
    }

\title{\boldmath Holographic timelike complexity for de Sitter}
\author[a]{Arpan Bhattacharyya}
\author[b]{, Md Khalid Hossain}
\author[b]{, Sanhita Parihar}
\author[b]{, and Shubho R. Roy}
\affiliation[a]{Department of Physics, Indian Institute of Technology Gandhinagar, Palaj, Gujarat 382355, India}
\affiliation[b]{Department of Physics, Indian Institute of Technology, Hyderabad, Kandi, Sangareddy, Telengana 502285, India}

\emailAdd{abhattacharyya@iitgn.ac.in, mdkhalid.ac@gmail.com, sanhita.hepth@gmail.com, roy.shubho@gmail.com}

\abstract{We investigate the recent proposal of holographic volume complexity for timelike subregions \cite{Alishahiha:2025xml} in the framework of static patch holography for de Sitter spacetime. Using the stretched-horizon prescription, we compute the timelike subregion complexity as a function of the subregion duration for pure de Sitter and Schwarzschild de Sitter geometries. In pure de Sitter spacetime, the timelike subregion complexity displays exponential growth for short durations, and hyperfast growth near a maximal duration, paralleling the features of spacelike volume complexity \cite{Jorstad:2022mls}. For Schwarzschild de Sitter, when the stretched horizon is near the cosmological horizon, the behavior broadly remains similar to pure de Sitter. However, when the stretched horizon is near the black hole horizon, the hyperfast growth for long durations is replaced by nonlinear growth regime. Along the way, we also compute the corresponding timelike holographic entanglement entropy for de Sitter and Schwarzschild de Sitter.}

\begin{document}
\maketitle
\flushbottom

\section{Introduction \& Summary}

Investigations into the dynamical properties of black holes and quantum fields in black-hole backgrounds during the 1960s and 1970s, in the pioneering works \cite{Bekenstein:2008smd,Hawking:1975vcx}, ultimately led to the formulation of the holographic principle \cite{tHooft:1993dmi,Susskind:1994vu} for quantum gravity, which states that a (quantum) gravitational theory in $(d+1)$ spacetime dimensions can be equivalently described by a nongravitational quantum theory living on its $d$-dimensional boundary. It was further generalized to a background independent covariant form via an entropy bound applicable to arbitrary spacetimes through the invocation of light-sheets \cite{Bousso:1999xy,Bousso:1999cb}. The holographic principle found its first concrete realization through the AdS/CFT correspondence arising in string theory \cite{Maldacena:1997re,Gubser:1998bc,Witten:1998qj,Aharony:1999ti}, which states that quantum gravity in asymptotically Anti de Sitter (AdS) spacetime is dual to a conformal field theory defined on its boundary. We now understand that every large $N$ factorizable conformal field theory provides a nonperturbative formulation of quantum gravity with asymptotically Anti de Sitter boundary conditions whereby classical spacetime and gravity are emergent \cite{El-Showk:2011yvt}. More recently, holographic descriptions of quantum gravity have also been proposed for asymptotically flat spacetimes, notable examples include the celestial holography program \cite{Pasterski:2016qvg, Pasterski:2021raf} and the Carrollian holography framework \cite{Duval:2014uva, Bagchi:2023cen, Ruzziconi:2026bix}, both of which exploit the symmetry structure of null infinity to formulate flat-space holography, and the older well-studied alternative, namely the M-theory/BFSS matrix model correspondence \cite{Banks:1996vh}, which casts flat space quantum gravity (M-theory) as a matrix model. Despite these remarkable successes, the holographic description of spacetimes with a positive cosmological constant remains far less understood. This issue is particularly important because observations indicate that the present Universe is undergoing accelerated expansion, while the inflationary epoch is well approximated by a quasi-de Sitter geometry \cite{SupernovaSearchTeam:1998fmf,SupernovaCosmologyProject:1998vns}. Consequently, de Sitter (dS) spacetime provides the simplest theoretical model for understanding quantum gravity in cosmological settings. De Sitter spacetime possesses a cosmological horizon, with an associated Gibbons-Hawking temperature and entropy \cite{Gibbons:1977mu},
\begin{equation}
S_{\rm dS}=\frac{A_{\rm H}}{4G},
\end{equation}
perhaps indicating the existence of a finite number of microscopic degrees of freedom. De Sitter spacetime has spacelike conformal boundaries, unlike AdS which has timelike conformal boundary, and observers in dS have access only to a finite region bounded by cosmological horizons. These features make the formulation of a holographic dual fundamentally different and substantially more challenging than in asymptotically AdS spacetimes or even flat spacetimes. Over the past two decades, several proposals have been put forward to establish a holographic description of de Sitter spacetime. Although none has yet attained the level of precision and completeness of the AdS/CFT correspondence, these approaches have significantly advanced our understanding of quantum gravity in spacetimes with a positive cosmological constant. Reviews of these developments can be found in references~\cite{Spradlin:2001pw,Anninos:2012qw,Galante:2023uyf}. The earliest formulation of physics in de Sitter spacetime based on holography is the $\text{dS/CFT}$ correspondence, which was based on symmetry arguments in \cite{Strominger:2001pn, Witten:2001kn} and further developed in \cite{Maldacena:2002vr}. In this picture, the wave functional, $\Psi_{\text{dS}}$, of the asymptotically dS$_{d+1}$ spacetime with boundary field configurations $\phi_0$ at future infinity, is identified with the partition function $Z_{\text{CFT}}$ of a CFT$_d$ deformed by sources $\phi_0$, expressed as $\Psi_{\text{dS}}[\phi_0] = Z_{\text{CFT}}[\phi_0]$. The chief advantage of $\text{dS/CFT}$ is that the isometry group $SO(d+1,1)$ of de Sitter space acts naturally as the global conformal symmetry group of the boundary CFT, allowing one to leverage the well-established mathematical apparatus of conformal field theories. However, the proposal suffers from critical drawbacks: the dual CFT is fundamentally non-unitary, exhibiting complex operator scaling dimensions and the hologram resides at spacelike future infinity $\mathcal{I}^+$, making it causally inaccessible to any localized interior observer. In contrast with the well established AdS-CFT duality where the equivalence is between two partition functions, here the equivalence is between a single state (wave functional) and a CFT partition function. One would rather like to obtain an answer to the question whether there exists a strictly Lorentzian and unitary conformal field theory in $d$-dimensions which describes the full Hilbert space (as opposed to a single wave functional) of asymptotically de Sitter (dS$_{d+1}$) quantum gravity (see \cite{Chowdhury:2021nxw,Chakraborty:2023yed} for some recent efforts). Efforts to embed de Sitter in string theory has also not been successful \cite{Dine:2020vmr}, beginning with powerful no-go theorems, e.g. \cite{Maldacena:2000mw, Robbins:2004hx, Green:2011cn, Kutasov:2015eba, Sethi:2017phn, Russo:2019fnk}  and culminating with the swampland conjecture(s) \cite{Obied:2018sgi, Ooguri:2018wrx}, which posit that quantum theories of gravity do not admit stable or metastable de Sitter space at all. These persistent difficulties in obtain a holographic description of global de Sitter spacetime, coupled with the well established tenets of (cosmological) horizon thermodynamics in de Sitter spacetime has motivated the formulation of alternative proposals for dS holography, such as the \emph{static patch holography} \cite{Susskind:2021omt, Susskind:2021esx} based on the physical experience of an observer in de Sitter whose access is limited to the a static patch of dS instead of the entire global dS. Rather than localizing the microscopic dual degrees of freedom to future infinity, this approach places the holographic boundary directly on a ``stretched horizon" placed just inside the cosmological horizon (a Planck distance inside the cosmological horizon). For the sake of completeness, we also mention the work \cite{Anninos:2011af}, where an alternative formulation of static patch holography was put forward, where the boundary dual theory was localized on the observer's worldline and reproduced the bulk correlation function underlying conformal symmetries of the worldline. A major breakthrough in realizing a concrete quantum mechanical dual for de Sitter space came through the Double-Scaled Sachdev-Ye-Kitaev (DSSYK) model, explored in \cite{Susskind:2022bia} as well as in \cite{Narovlansky:2023lfz}. By working with the DSSYK model, a system of $N$ randomly interacting Majorana fermions with $p$-body interactions in the double-scaling limit around the top edge of its energy spectrum, the microscopic chord diagram combinatorics match the chord algebra of Jackiw-Teitelboim ($\text{JT}$) gravity and 2+1-dimensional de Sitter space. This framework provides an explicitly solvable, finite-dimensional quantum mechanical Hamiltonian governed by $q$-deformed harmonic oscillator algebras that reproduces hyper-fast scrambling \cite{Susskind:2021esx}, finite entropy \cite{Narovlansky:2023lfz}, and two-point correlators along an observer's worldline \cite{Narovlansky:2023lfz,Verlinde:2024zrh}. However, the primary downside of this setup is that the explicit duality is currently restricted to lower-dimensional toy models ($\text{dS}_2$ and $\text{dS}_3$), and fine-tuning the double-scaling parameters to achieve a wide hierarchy between the Planck scale and the cosmological horizon scale requires delicate limit manipulations. 
Finally, originating from the study of solvable irrelevant deformations in two-dimensional field theories, the $T\bar{T} + \Lambda_2$ deformation framework proposed in \cite{Lewkowycz:2019xse} and expanded in \cite{Coleman:2021nor}, constructs de Sitter static patch by deforming an existing $\text{AdS/CFT}$ dual. Beginning with a 2d CFT dual to a BTZ black hole, one applies a quadratic stress-tensor $T\bar{T}$ deformation, which holographically corresponds to moving the asymptotic Dirichlet boundary of $\text{AdS}$ inward all the way to a stretched horizon close to the black hole horizon. By modifying the flow equation with a positive energy density term ($\Lambda_2 > 0$) thereby switching the geometric signature from AdS to dS, one obtains a stretched horizon enclosing a $\text{dS}_3$ static patch. The trade-off is that $T\bar{T}$ is an irrelevant operator that renders the boundary theory non-local at high energies (exhibiting Hagedorn-type behavior or complex energy eigenvalues), and extending these stress-tensor deformations to higher dimensions ($\text{dS}_{d+1}$ for $d > 2$) presents significant mathematical obstacles.
    
In holography, quantum information tools such as entanglement entropy (EE) and complexity have been proven to be extremely useful in providing insights into the underlying nature of holography. A major breakthrough in this direction was the Ryu-Takanayagi (RT) proposal \cite{Ryu:2006bv,Ryu:2006ef}. This proposal has significantly advanced our understanding of how the quantum information of the boundary theory (measured by the EE) is encoded into the bulk geometry of AdS spacetime and vice versa \cite{Maldacena:2013xja,Faulkner:2013ica}. While the RT proposal of holographic entanglement, and its quantum extensions such as the quantum extremal surface (QES) \cite{Engelhardt:2014gca} has been remarkably successful in explaining aspects of Hawking radiation from an evaporating AdS black hole by leading the way towards the island proposal \cite{Penington:2019npb, Almheiri:2019qdq, Penington:2019kki}, it has some limitations. It fails to capture the linear growth of the Einstein-Rosen bridge for eternal AdS black holes. This shortcoming motivated the introduction of another quantum information measure, namely quantum computational complexity. The holographic dual of quantum computational complexity of boundary theory is conjectured to be the maximal volume of a spacelike hypersurface that successfully captures the growth of the Einstein-Rosen bridge \cite{Susskind:2014rva}. Consequently, holographic complexity has emerged as a powerful tool to study black hole interiors. Nowadays, there are several different gravitational observables which are conjectured as holographic duals of quantum computational complexity of boundary field theory; each captures the universal properties of the interior of black holes \cite{Belin:2021bga, Belin:2022xmt}. The success of these quantum information-based probes of boundary field theory towards the underlying holography in asymptotically AdS has motivated us to study them in the context of dS holography as well. There were various approaches towards extending the RT proposal to de Sitter spacetime. One strategy involves realizing the de Sitter boundary within an AdS spacetime \cite{Hawking:2000da}. Others are specific to each dS holography proposal\footnote{For alternative approaches, readers are referred to \cite{Arias:2019pzy,Arias:2019zug,Arenas-Henriquez:2022pyh,Maulik:2026vmj}.}. In the dS/CFT correspondence, timelike codimension two extremal surfaces connecting future and past null infinities are considered \cite{Narayan:2017xca,Narayan:2022afv, Narayan:2023zen} . For the dS static patch holography, there are two different holographic proposals for EE, namely the monolayer and the bilayer \cite{Shaghoulian:2022fop}. The monolayer proposal \cite{Susskind:2021esx} states that the EE between a subregion and its complement situated at the two stretched horizons is given by the area of a extremal surface which is homologous to the subregions and extends between the two stretched horizons. While the bilayer proposal \cite{Shaghoulian:2021cef} conjectures that the EE of the same is given by the area of an extremal surface which is homologous to the subregions and extends in both the exterior and the two interiors. 

More recently, the concept of EE has been extended to timelike subregions, leading to the notion of timelike entanglement entropy (TEE), which characterizes the entanglement between temporal segments rather than spatial regions \cite{Doi:2022iyj,Doi:2023zaf}, first proposed in the context of dS/CFT correspondence \cite{Strominger:2001pn}, to generalize the notion of entanglement between time segment in the absence of timelike asymptotic boundary. Holographically, TEE is given by the combined area of spacelike and timelike extremal surfaces homologous to the boundary time subregion. Unlike its purely spacelike counterpart, the holographic timelike entanglement entropy is in general complex valued, and the imaginary part has been interpreted as \textit{emergent time}. Recently, a geometric interpretation of the imaginary part was given by relating it to complex extremal surfaces probing the black hole interior \cite{Heller:2024whi}. On the field theory side, it is identified with \textit{pseudo entropy} which is defined as the von Neumann entropy corresponding to the transition matrix between two different states \cite{Doi:2023zaf}. Soon after, the holographic TEE was studied in the specific case of $AdS_{3}/CFT_{2}$, and the extremization prescription of the spacelike and timelike areas is discussed in detail \cite{Li:2022tsv}. In the past few years, holographic TEE has been examined across a wide range of settings: deformed and finite-temperature CFTs, black hole, braneworld backgrounds, and de Sitter geometries \cite{Chu:2023zah,Jiang:2023ffu,Afrasiar:2024ldn,Afrasiar:2024lsi,Chu:2025sjv,Nunez:2025gxq,Narayan:2026wzp}. Its behavior has been understood from the perspective of renormalization group flow \cite{Jiang:2023loq}, where the imaginary part is shown to track the flow between fixed points. Holographic TEE has also found application as a diagnostic tool for probing bulk reconstruction, offering a new lens to understand the encoding of causal and temporal structure of the bulk in boundary data \cite{Das:2023yyl}. Furthermore, an entanglement first law has been derived for TEE, extending the linearized-Einstein-equation derivation to the timelike case and tying the emergent-time data to bulk gravitational dynamics perturbatively \cite{Li:2025tud}. 
Motivated by these developments, it is natural to ask whether quantum complexity admits a similar extension to timelike subregions. In a recent work \cite{Alishahiha:2025xml}, a proposal has been put forward to generalize the holographic ``Complexity=Volume" proposal to boundary time intervals. Also see \cite{Prihadi:2026nua,Pal:2026ysc,Li:2026bof} for some follow-up works. The construction begins by extremizing the relevant area functional for a timelike boundary subregion, which yields two distinct branches of extremal surfaces a spacelike branch and a timelike branch. These two branches together form the complete extremal surface configuration bounding the region whose volume is defined as the timelike subregion complexity. A striking result of this construction is that the resulting subregion complexity remains purely real, unlike the analogous timelike entanglement entropy calculation, which acquires an imaginary part. This reality persists even when the analysis is extended to AdS black brane geometries, where the extremal surfaces may either stay outside or penetrate the horizon. Also, the computed complexity shows the same universal UV divergences as the spacelike case. In the present work,
we investigate the notion of timelike subregion complexity \cite{Alishahiha:2025xml} for de Sitter spacetime, adopting the static patch holography framework for our analysis, following up on the work on holographic complexity in de Sitter spacetime \cite{Jorstad:2022mls}. In particular, we investigate whether this construction continues to yield a real and causally well-defined notion of complexity in the presence of a cosmological horizon, or whether the intrinsic compactness of de Sitter space and the absence of a genuine asymptotic timelike boundary introduce new subtleties that are not present in the AdS setting.\\

The paper is organized as follows. In Section~(\ref{C_spacelike}) we revisit the analysis of the full volume complexity at fixed boundary times, i.e., spacelike sections of the boundary performed numerically, and highlight the salient features. At early times, full volume complexity displays exponential growth with a characteristic growth rate constant independent of spacetime dimensionality, $\mathcal{C}_V\sim e^{T}$. We also note the existence of a finite maximal/critical time $T_{\textrm{max}}$ whereby complexity diverges, i.e., the \emph{hyperfast growth} phenomenon. We also numerically confirm the manner in which the complexity diverges while approaching the critical time, i.e., a power law divergence with exponent $(d-1)$, where $d$ is the number of spatial dimensions, $ \mathcal{C}_V \sim \left(T_{\textrm{max}}-T\right)^{-(d-1)}$. This section provides context and serves as an important reference point for the ensuing analysis of timelike subregion complexity. In Section~(\ref{d=2}) we first consider the three-dimensional de Sitter spacetime ($dS_3$), since it is the simplest case and allows for an analytical treatment. We consider a subregion of the stretched horizon extended in the time direction. We first review the prescription \cite{Alishahiha:2025xml} for timelike subregion volume complexity and adapt it to the de Sitter case for static patch holography. Then we work out the spacelike extremal curves anchored at the past and future extremities of the subregion and the associated timelike extremal curve. Subsequently, we compute the timelike entanglement entropy and the timelike subregion volume complexity and study their behaviour. We find that the timelike entanglement entropy has both real and imaginary parts, both of which remain finite and positive in this limit. The timelike subregion volume complexity exhibits the same traits as the spacelike volume complexity, namely, an early-time exponential growth followed by hyperfast growth at a finite maximal time, $T_{\textrm{max}}$, with a logarithmic divergence 
\begin{equation}
   C^{\textrm{late}}_{s}\sim \ln\left(T_{\textrm{max}}-T\right)^{-1}. 
\end{equation}
 This provides a clear understanding of the main features before studying higher-dimensional de Sitter spacetimes, where analytical results are generally difficult to obtain. In Section~(\ref{sec:dS_d}) we present a numerical study of timelike entanglement entropy and timelike complexity for a general $(d+1)$ dimensional de Sitter spacetime. We use the analytic results found in Section ~(\ref{d=2}) for three-dimensional de Sitter spacetime to validate our numerical computations and thereby lend confidence to the numerical results found for higher dimensions. In the numerical study, we find that timelike complexity in higher-dimensional spacetime shares the same qualitative features as have been observed in three-dimensional de Sitter spacetime. It is a monotonically increasing function of the subregion duration, and there exists a finite critical time (duration) $T_{\text{max}}$, which is independent of the dimension of spacetime, whereby the timelike complexity diverges. It exhibits a hyperfast growth near the critical duration, with a power law dependence over boundary subregion duration $T$ as shown in Fig.~(\ref{fig:logCvslogT_latetime}),
\begin{align}
    C_{s}^{\text{late}}\sim \frac{1}{(T_{\text{max}}-T)^{d-2}},
\end{align}
with an dimension dependent exponent $(d-2)$. For subregion durations significantly shorter than $T_{\textrm{max}}$ (early times), the timelike subregion complexity displays exponential growth as shown in Fig.~(\ref{fig:logCvsT_earlytime}) with the growth rate being independent of spacetime dimensionality. In addition to that, an explicit dependence of timelike complexity over subregion time for $d=3,4,5$ is shown in Fig.~(\ref{fig:complexity_num}). 
Next, in Section~(\ref{sec:SdS_BH}), we undertake the study of the timelike complexity of Schwarzschild de Sitter black holes. In the de Sitter black hole case, analytical treatment was not feasible even for the case of the lowest nontrivial dimensions, namely $d=3$, and we had to proceed numerically exclusively. The Schwarzschild de Sitter geometry allows various possibilities for the placement of the stretched horizon, and we consider two different positions of the stretched horizon as representative cases. When the stretched horizon lies near the cosmological horizon, the geometry has much similarity to the empty de Sitter, and this is the case we first study in Section~(\ref{subsec: case1}). As a result of this study, we find that in this particular configuration of stretched horizon, even the Schwarzschild black hole exhibits a late-time hyperfast growth as shown in Fig.~(\ref{fig:logCvsT_latetime_BH1}), and an exponential growth for short subregion durations. In Section~(\ref{subsec: case2}), we consider the interesting case where the stretched horizon is near the black hole horizon, as this configuration is rather different in comparison to the empty de Sitter. In this case, our findings suggest that even though timelike complexity exhibits an exponential growth in early time at late time, there is no hyperfast growth phenomenon, i.e., the timelike complexity does not blow up at a finite time. Instead, the initial exponential growth is simply softened to a power law increase with the boundary subregion duration. Finally, in Section~(\ref{Conclusions}), we discuss our results in light of various results obtained in the recent literature and provide an outlook for future investigations. In the appendices, we provide some supporting calculations. In Appendix~(\ref{A1}) we back up the numerical calculations of Section~(\ref{sec:dS_d}) by performing the numerical analysis in Eddington-Finkelstein coordinates. In Appendix ~(\ref{A2}) we provide an analytical derivation of the power-law growth of timelike subregion volume complexity for SdS black holes in case 2 mentioned in Section~(\ref{subsec: case2}), where the stretched horizon is placed right outside the black hole horizon.

\section{Complexity for spacelike subregions}\label{C_spacelike}

\begin{figure}[htb!]
    \centering
    \includegraphics[width=0.6\linewidth]{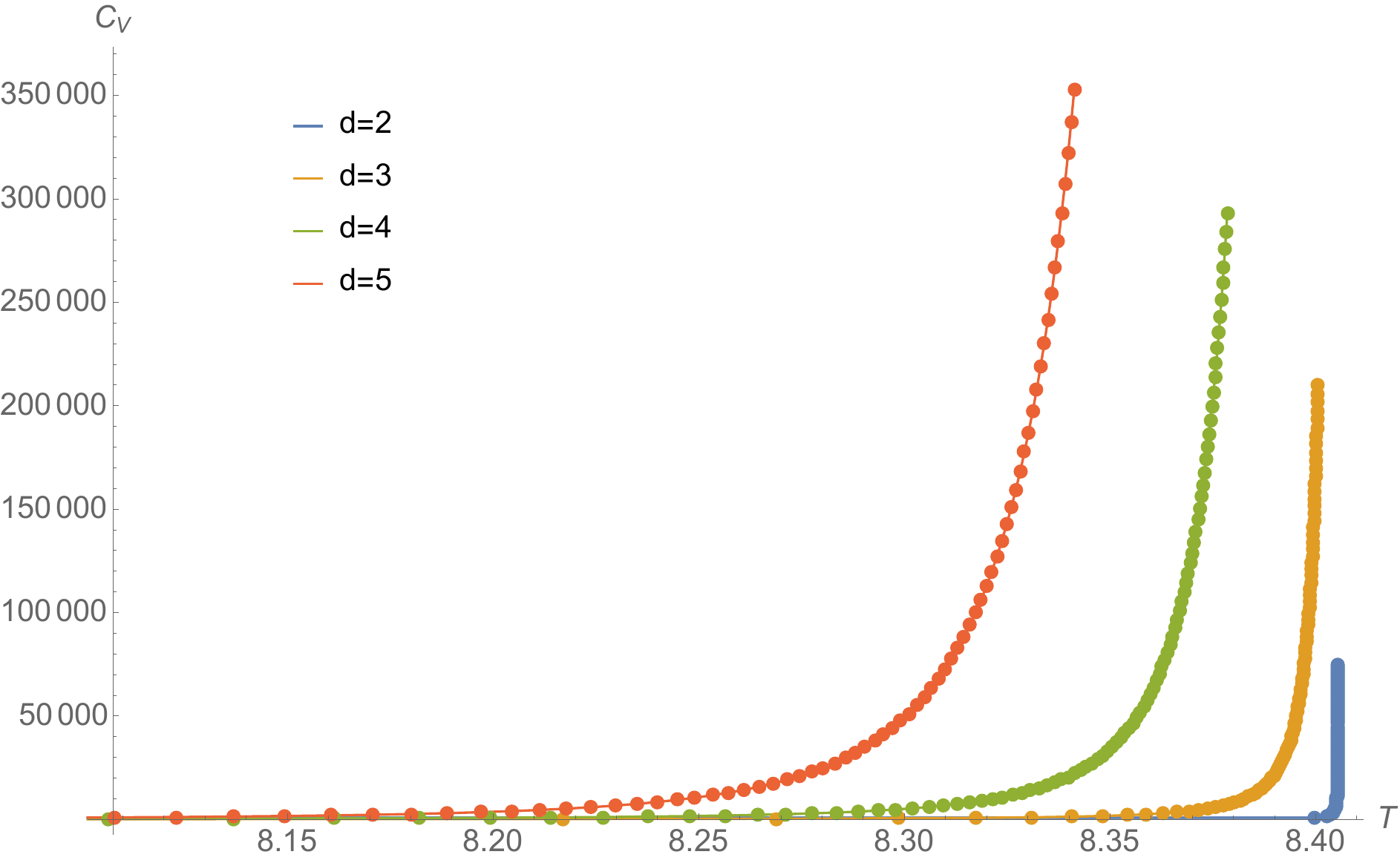}
    \caption{Plot of $C_V$ for pure dS as a function of the boundary time $T$ for stretched horizon cutoff parameter $\epsilon=10^{-7}$ for diverse dimensions $d=2,3,4,5$.}
    \label{CVvTs}
\end{figure}

Right at the outset, we revisit the analysis of the holographic volume complexity for a spacelike subregion in the context of the static patch holography along the lines of \cite{Jorstad:2022mls} to highlight some of the features of this volume complexity in de Sitter spacetime, including some of the omitted details (e.g., the exponential time-dependence at early times). This exercise will facilitate a comparison of the complexity of timelike subregions obtained in the later section, especially the time dependencies at both early and late times (near a critical time). 
To this end, we switch to outgoing null coordinates $u$ in lieu of the Schwarzschild time $t$, whereby the metric assumes the form,
\begin{equation*}
    ds^2= -f(r)\,du^2 -2\,du\,dr + r^2 d\Omega_{d-1}^2
\end{equation*}
where $f(r)=1-r^2$.  We have set the de Sitter radius to unity $l=1\,.$ Owing to spherical symmetry, the maximal volume hypersurface will be of the form $r=r(u)$. The volume functional to be maximized is,
\begin{equation*}
    \mathcal{V}=\int du\,L(r,\dot{r}),\,L=V_{S^{d-1}}\,r^{d-1}\sqrt{-f(r)-2\dot{r}} 
\end{equation*}
where $\dot{r}=\frac{dr}{du}$. Since the volume functional is $u$-independent, there exists a first integral, namely
\begin{equation*}
    H \equiv \frac{\partial L}{\partial\dot{r}} \dot{r}-L = V_{S^{d-1}}\,r^{d-1} \frac{f(r)+\dot{r} }{\sqrt{-f(r)-2\dot{r}}}\,.
\end{equation*}
At the point on the extremal volume slice where $r$ hits a maximum i.e $\dot{r}=0,$ we have
\begin{equation*}
    H= -V_{S^{d-1}}\,r_{\textrm{max}}^{d-1} \sqrt{-f(r_{\textrm{max}})}\,.
\end{equation*}
Evidently, this point is attained outside the horizon $r>1$ for which $f(r)<0$. In terms of the turning point one has,
\begin{equation*}
    \dot{r} = -F(r),\quad F(r)\equiv \sqrt{f(r)+\frac{\left|f(r_{\textrm{max}})\right| r^{2(d-1)}_{\textrm{max}}}{r^{2(d-1)}}}\left[\sqrt{f(r)+\frac{\left|f(r_{\textrm{max}})\right| r^{2(d-1)}_{\textrm{max}}}{r^{2(d-1)}}}\pm \sqrt{\frac{\left|f(r_{\textrm{max}})\right| r^{2(d-1)}_{\textrm{max}}}{r^{2(d-1)}}} \right]\,.
\end{equation*}
The upper sign corresponds to the global maxima of the volume as the quantity inside the square root in the integrand of the volume functional is larger. The closed-form solution, with the boundary condition that the curve intersects the stretched horizon at $t=T,r=1-\epsilon$, is
\begin{equation}
    u(r) =T-\frac{1}{2}\ln{\frac{2-\epsilon}{\epsilon}} - \int_{1-\epsilon}^r \frac{dx}{F(x)}\,. \label{spacelike extremal in outgoing EF}
\end{equation}
At the turning point, owing to the symmetry of the maximal volume surface, it lies on the curve $t=0$, so $u(r_{\textrm{max}})=-\tilde{r}(r_{\textrm{max}})$. Explicitly,
\begin{equation*}
    u(r_{\textrm{max}}) = \frac{1}{2}\ln\left(\frac{r_{\textrm{max}}-1}{r_{\textrm{max}}+1}\right).
\end{equation*}
Plugging this back in the solution \eqref{spacelike extremal in outgoing EF} one obtains a condition relating the boundary time $T$ and the turning point coordinate $r_{\textrm{max}}$ (for a fixed $\epsilon$),
\begin{equation}
T=\frac{1}{2}\ln\left(\frac{r_{\textrm{max}}-1}{r_{\textrm{max}}+1}\right)+\int_{1-\epsilon}^{r_{\textrm{max}}} \frac{dx}{F(x)}+\frac{1}{2}\ln{\frac{2-\epsilon}{\epsilon}}\,. \label{T in terms of epsilon and rmax}
\end{equation}
Evidently, there is an upper limit on $T$ set by the value corresponding to $r_{\textrm{max}}\rightarrow \infty\,.$ 
\begin{equation}
    T_{\textrm{max}}=  \frac{1}{2}\ln{\frac{2-\epsilon}{\epsilon}}\,. \label{Maximum time for spacelike}
\end{equation}
At $T_{\textrm{max}},$, the complexity blows up, see below. Then the volume complexity for spacelike subregion is,
\begin{equation}
    \mathcal{C}_V=\frac{1}{G_N} \mathcal{V}= \frac{2 V_{S^{d-1}}}{G_N} \int_{1-\epsilon}^{r_{\textrm{max}}} dr\,r^{d-1}\frac{\sqrt{-f(r)-2\dot{r}}}{-\dot{r}}=\frac{2V_{S^{d-1}}}{G_N} \int_{1-\epsilon}^{r_\textrm{max}} dr\,r^{d-1}\frac{\sqrt{2F(r)-f(r)}}{F(r)}\label{Cv spacelike}
\end{equation}
obtained as function of the turning point $r_{\textrm{max}}$ and the cutoff parameter $\epsilon$. Subsequently, eliminating $r_{\textrm{max}}$ from \eqref{T in terms of epsilon and rmax} and \eqref{Cv spacelike} allows us to (numerically) extract the dependence of volume complexity on the boundary time. The numerical plots of the $\mathcal{C}_V$ as a function of the boundary time $T$ for a fixed value of the stretched horizon cutoff  $\epsilon=10^{-7}$ are shown here in Fig.~(\ref{CVvTs}).
From the plots, the following features of the full boundary (volume) complexity, $\mathcal{C}_V$ can be extracted:
\begin{itemize}
    \item For a fixed value of the cutoff $\epsilon$, the complexity grows exponentially for early times, \begin{equation}
        \mathcal{C}_V(T)\sim e^{T}.
    \end{equation} 
    This can be surmised from the plot of $\ln{\mathcal{C}_V}$ vs. $T$ for ``early times", which is linear for values of $T$ appreciably shorter\footnote{To obtain an idea about how long the exponential growth of volume complexity persists in comparison to the maximal/critical time $t_{\textrm{max}}$ whereby complexity blows up, consider the nominal case when the stretched horizon cutoff is $\epsilon=10^{-7}$. In this case, the exponential growth carries on till $T\approx 6.5$, while $T_\textrm{max}\approx8.4$ in dS radius units.}  than $T_{\textrm{max}}$ with slope unity as shown in Fig.~(\ref{logCVvTs}).
    \begin{figure}[h]
    \centering
    \includegraphics[width=0.60\linewidth]{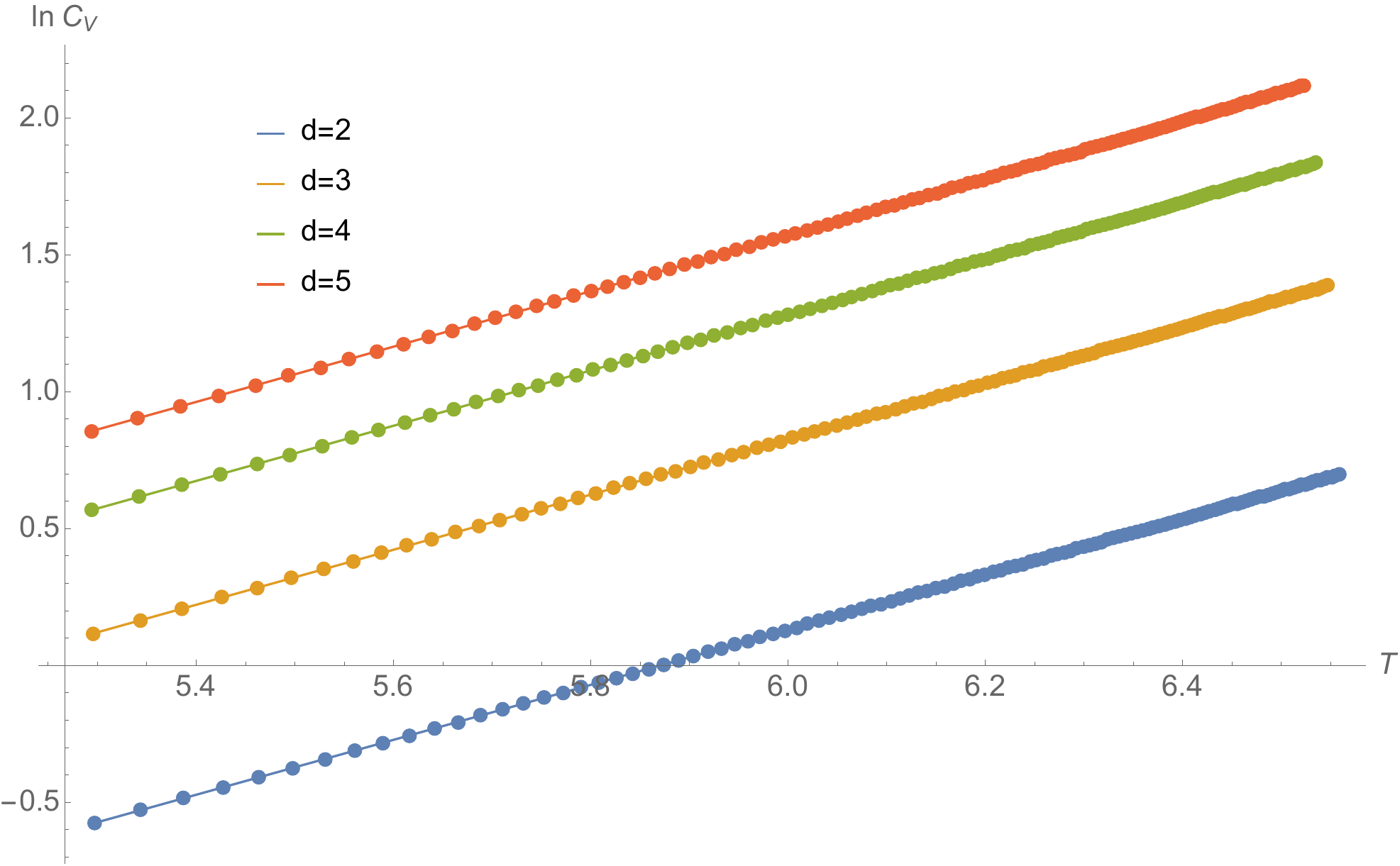}
    \caption{Plot of $\ln\,C_V$ for pure dS as a function of the boundary time $T$ at ``early times" $T<6.5$  for stretched horizon cutoff parameter $\epsilon=10^{-7}$ for diverse dimensions $d=2,3,4,5$: all are straight lines with slope unity. Note that the critical time is $T_{\textrm{max}}\approx8.4$.}
    \label{logCVvTs}
\end{figure}
This is independent of the dimension, as has been checked for $d=2,3,4,5$. Contrast this with volume complexity growth in global de Sitter space, where the volume of the spatial sections grows exponentially (inflate) with time \cite{Parihar:2026rce}, and owing to extensivity, so does the volume complexity, $C_V \sim e^{(d-1)T}$. Here we are working in the static patch where the boundary CFT lives on a static background, and hence this exponential growth must be accounted for by a completely different mechanism, e.g., the explicit time-dependence of the marginal couplings. We would like to remark that this early time exponential growth is much akin to Krylov complexity, as has been conjectured \cite{Parker:2018yvk} for chaotic quantum systems, and recorded in the literature for doubled-scaled SYK model \cite{2023JHEP...08..099B, Anegawa:2024yia} and also in 2d CFTs \cite{Dymarsky:2021bjq}, in free and interacting QFTs \cite{Avdoshkin:2022xuw,Camargo:2022rnt}.
 \begin{figure}[htb!]
    \centering
    \includegraphics[width=0.6
     \linewidth]{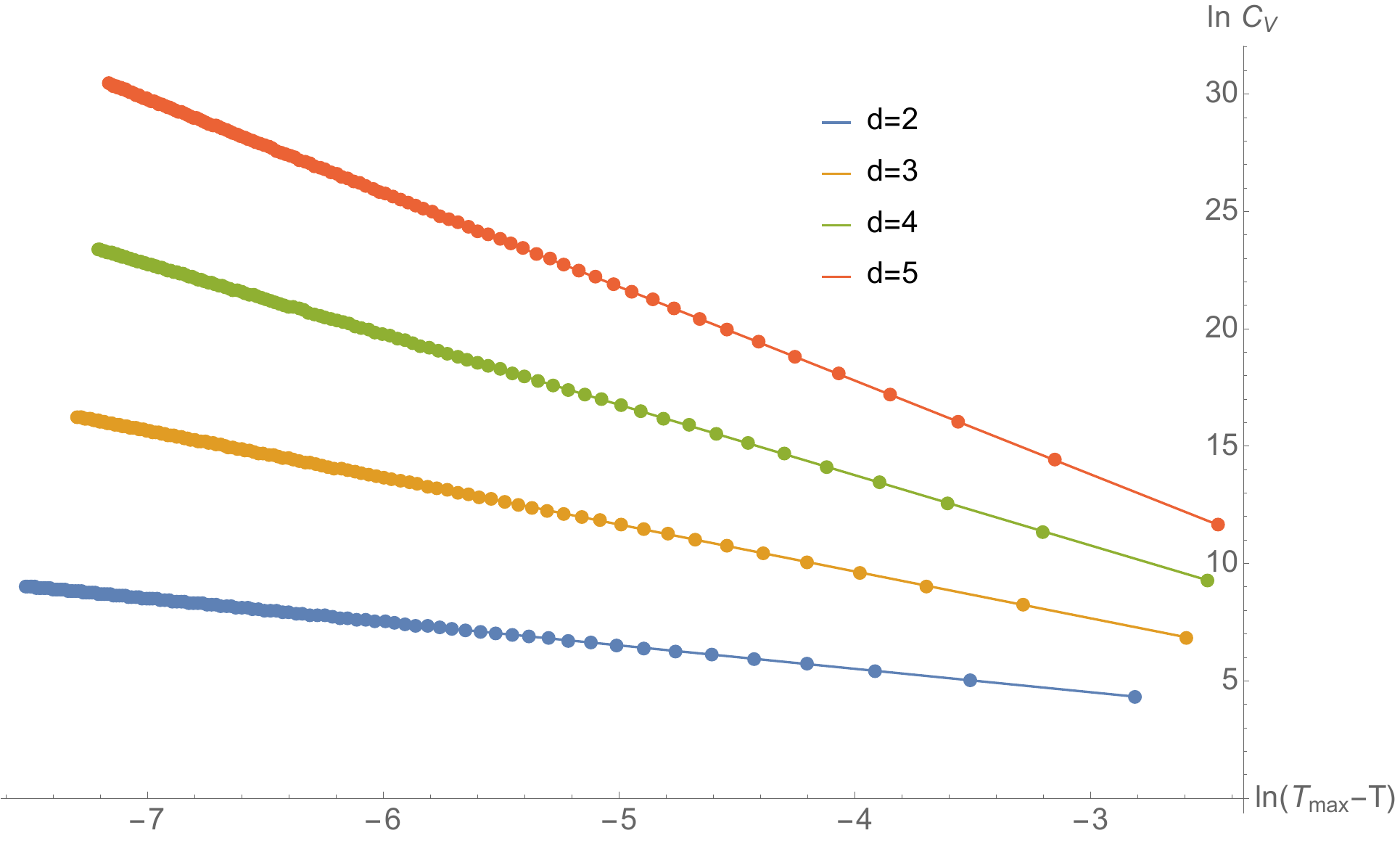}
    \caption{Plot of $\ln\,C_V$ for pure dS as a function of $\ln(T_{\textrm{max}}-T)$ for ``late times" $T$, i.e. close to $T_{\textrm{max}}\sim 8.4$ with the stretched horizon cutoff parameter set to $\epsilon=10^{-7}$ for diverse dimensions $d=2,3,4,5$: all are straight lines with slope $-(d-1)$. This represents hyperfast growth.}
    \label{logCVvlogTs}
\end{figure}

    \item For fixed cutoff, at ``late times", namely close to the critical or maximal time $T_{\textrm{max}}$, the complexity exhibits \emph{hyperfast growth}, namely complexity diverges at a finite critical time, $T_{\textrm{max}}$, as a power-law with exponent $(d-1)$:
    \begin{equation}
        \mathcal{C}_V \sim \frac{1}{(T_{\textrm{max}}-T)^{d-1}}\,.
    \end{equation}
    This can be surmised from the numerical plot of $\ln\mathcal{C}_V$ vs $\ln(T_{\textrm{max}}-T)$ in Fig.~(\ref{logCVvlogTs}), which are all straight lines with slopes equal to $-(d-1)$. 
       \item For fixed time $T$, the volume complexity increases monotonically with the stretched horizon cutoff $\epsilon$, refer to Fig.~(\ref{EpsCvvT}). Notably, no divergences are encountered in complexity as $\epsilon$ is sent to zero. However, the maximal or critical time $T_{max}$ diverges as $\epsilon$ is sent to zero, as evident from \eqref{T in terms of epsilon and rmax}. So, in this sense, $\epsilon$ appears to play the role of an large time IR regulator in the boundary theory.
        \begin{figure}[h]
    \centering
    \includegraphics[width=0.6
     \linewidth]{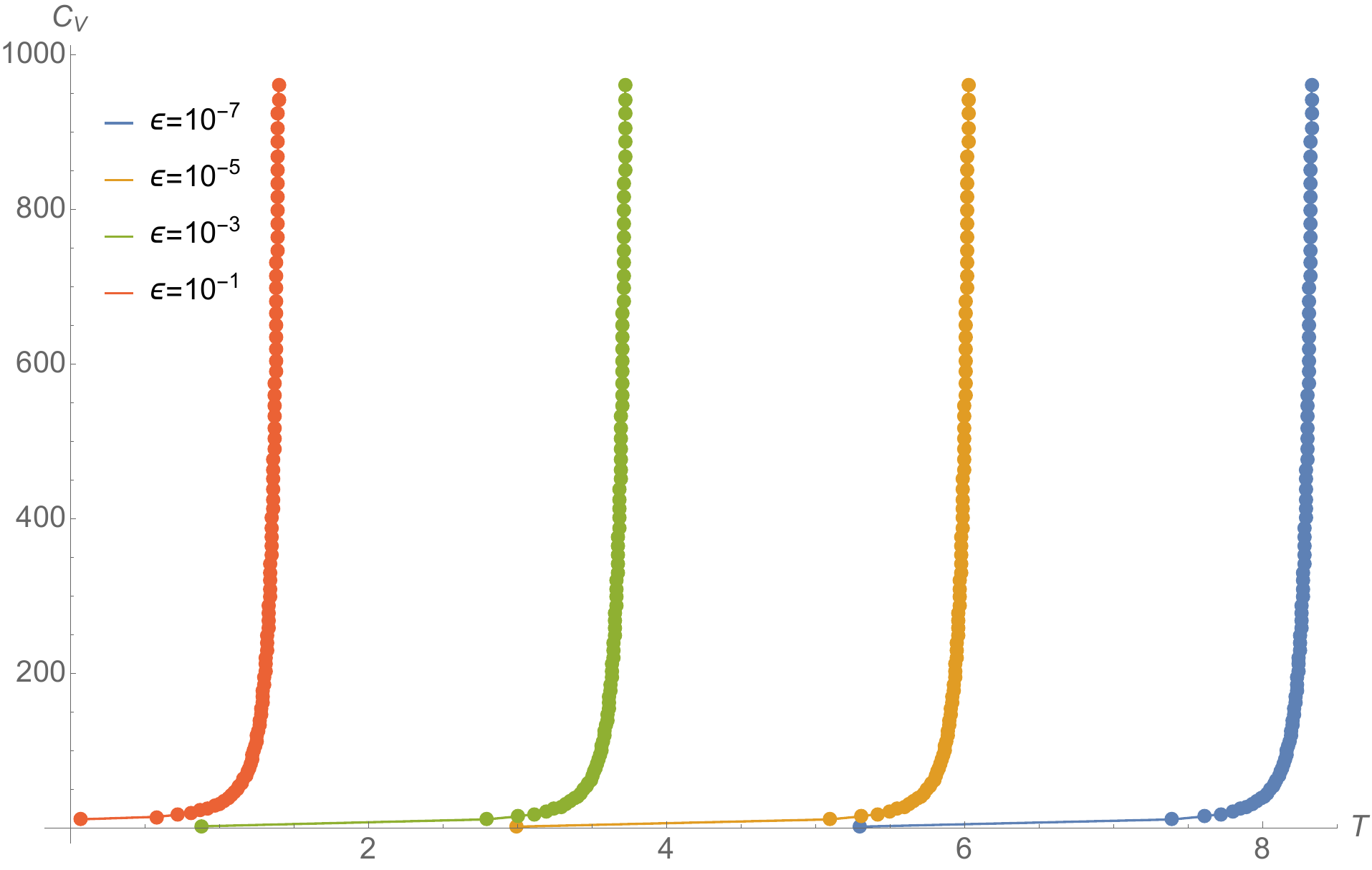}
    \caption{Comparison of volume complexity $C_V$ for pure dS with $d=3$ as a function of $T$ for diverse values of the stretched horizon cut off parameter $\epsilon$.}
    \label{EpsCvvT}
    \end{figure}
\end{itemize}

\section{Three dimensional de-Sitter Spacetime: Timelike extremal surface} \label{d=2}

In this section we work out the timelike subregion complexity for three dimensional pure de Sitter spacetime (dS$_3$). In three dimensions, the relevant computations can be performed analytically in Kruskal coordinates, unlike in higher dimensions, where a numerical treatment is generally required. First, we briefly review the dS$_3$ geometry in Schwarzschild and global Kruskal coordinates and spell out our conventions for defining local Eddington-Finkelstein coordinates for future reference. We set the de Sitter radius to unity, $L=1$. The static-patch metric is
\begin{equation}
ds^2=-(1-r^2)\,dt^2+\frac{dr^2}{1-r^2}+r^2d\phi^2.
\end{equation}

\begin{figure}[h]
\centering
\includegraphics[width=0.45\linewidth]{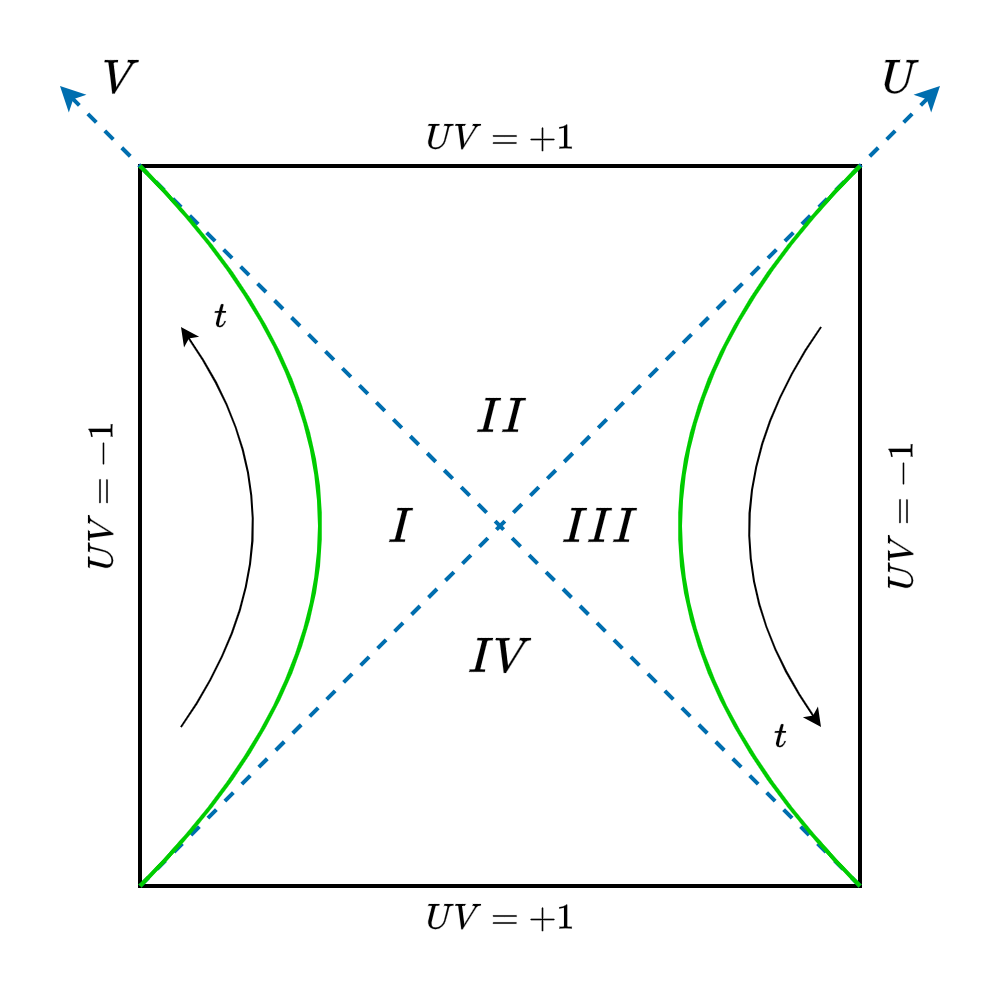}
\caption{Penrose diagram of de Sitter spacetime in global Kruskal coordinates. The cosmological horizons $U=0$ and $V=0$ divide the spacetime into four regions. The green curve denotes the stretched horizon, $r=1-\epsilon$, and the arrow indicates the direction of increasing Schwarzschild time in region $I$.}
\label{P1}
\end{figure}
In region $I$, where $r<1$, the tortoise coordinate is
\begin{equation}
\tilde r=\frac{1}{2}\ln\left(\frac{1+r}{1-r}\right),
\end{equation}
and the null Eddington--Finkelstein coordinates are defined by,
\begin{equation}
u=t-\tilde r\,,
\qquad
v=t+\tilde r\,.
\end{equation}
The corresponding global Kruskal coordinates are:
$U=-e^{-v}\,, V=e^u,$
which imply
\begin{equation}
UV=-e^{-2\tilde r},
\qquad
\frac{U}{V}=-e^{-2t}.
\end{equation}
The dS$_3$ metric in Kruskal coordinates becomes
\begin{equation}
ds^2=-\frac{4\,dUdV}{(1-UV)^2} + \frac{(1+UV)^2}{(1-UV)^2}\,d\phi^2,
\label{dS in Kruskal}
\end{equation}
where the Schwarzschild radial coordinate is related to the Kruskal coordinates by
$$r=\frac{1+UV}{1-UV}\,.$$.\
For $r>1$, the tortoise coordinate is defined as
\begin{equation}
\tilde r=\frac{1}{2}\ln\left(\frac{r+1}{r-1}\right).
\end{equation}
The transformations between the local Eddington--Finkelstein and global Kruskal coordinates in the four regions are summarized in Table~(\ref{tab:kruskal-map}).
\begin{table}[htb!]
\centering
\renewcommand{\arraystretch}{1.3}
\setlength{\tabcolsep}{10pt}
\begin{tabular}{ccc}
\hline
Region & $U$ & $V$ \tabularnewline
\hline
$I$ & $-e^{-v}$ & $e^u$ \tabularnewline
$II$ & $e^{-v}$ & $e^u$ \tabularnewline
$III$ & $e^{-v}$ & $-e^u$ \tabularnewline
$IV$ & $-e^{-v}$ & $-e^u$ \tabularnewline
\hline
\end{tabular}
\caption{Transformation from the local Eddington--Finkelstein coordinates $(u,v)$ to the global Kruskal coordinates $(U,V)$ in the four regions of dS$_3$.}
\label{tab:kruskal-map}
\end{table}
With these conventions, Schwarzschild time increases upward in region $I$ and downward in region $III$. The important geometric loci are summarized in Table~(\ref{tab}).

\begin{table}[h]
\centering
\renewcommand{\arraystretch}{1.35}
\begin{tabular}{ccc}
\hline
Geometric locus & Schwarzschild coordinate & Kruskal-coordinate condition \tabularnewline
\hline
North pole & $r=0$ & $UV=-1$ \tabularnewline
Cosmological horizon & $r=1$ & $UV=0$ \tabularnewline
Future infinity & $r\rightarrow\infty$ & $UV=1$ \tabularnewline
Static patch, region $I$ & $0<r<1$ & $-1<UV<0$ \tabularnewline
Stretched horizon & $r=1-\epsilon$ &
$\displaystyle UV=-\frac{\epsilon}{2-\epsilon}$ \tabularnewline
\hline
\end{tabular}
\caption{Important geometric loci of dS$_3$ in Schwarzschild and global Kruskal coordinates.}
\label{tab}
\end{table}

Before ending this section, note that, in the static-patch holography proposal, the dual field theory is supported on the timelike stretched horizon,
$r=1-\epsilon\,.$ We will be going to use this fact throughout the rest of the paper.

\subsection{Timelike Entanglement Entropy}

Recently, a holographic proposal for timelike entanglement entropy was put forward by Takayanagi \textit{et al.}~\cite{Doi:2022iyj,Doi:2023zaf}, extending the notion of entanglement entropy to timelike subsystems through timelike extremal surfaces in the context of AdS. Here, we compute the timelike entanglement entropy in dS. 
We consider a timelike subregion on the stretched horizon:  $-T<t<T$, $\phi,r = \text{constant}$. 
At $r = 1 - \epsilon$, the tortoise coordinate becomes
\begin{equation}
\tilde{r} = \frac{1}{2}\ln\!\left(\frac{1+r}{1-r}\right)
= \frac{1}{2}\ln\!\left(\frac{2 - \epsilon}{\epsilon}\right)\,,
\end{equation}
which gives
\begin{equation}
UV = -e^{-2\tilde{r}} = -\frac{\epsilon}{2 - \epsilon}\,.
\end{equation}
Evaluating at $t = T$, we also have
\begin{equation}
\frac{U}{V} = -e^{-2T}.
\end{equation}
Thus, the Kruskal coordinates satisfy
\begin{equation}
U = -\sqrt{\frac{\epsilon}{2 - \epsilon}}\, e^{-T}, 
\qquad
V = \sqrt{\frac{\epsilon}{2 - \epsilon}}\, e^{T}. \label{boundary}
\end{equation}
For later convenience, we introduce the notation
\begin{equation}\label{eq:U0_V0}
U^{0}=\sqrt{\frac{\epsilon}{2 - \epsilon}}\, e^{-T}, 
\qquad
V^{0}=\sqrt{\frac{\epsilon}{2 - \epsilon}}\, e^{T}.
\end{equation}
\begin{center}
 \begin{figure}[htb!]
    \centering
    \includegraphics[width=0.6\linewidth]{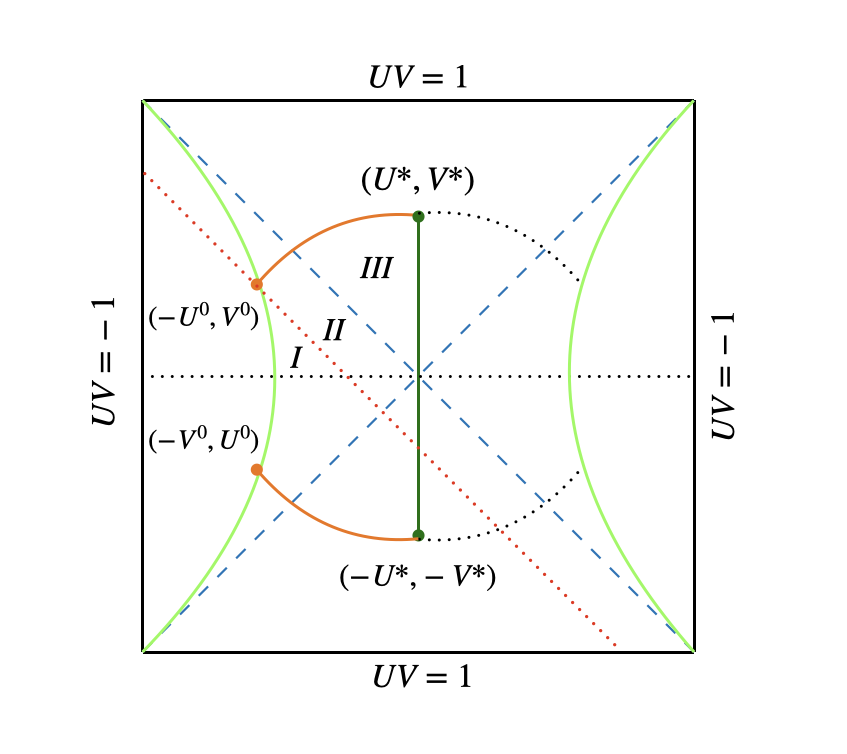}
    \caption{Penrose diagram of three-dimensional de Sitter spacetime, light green curves represent the stretched horizons, the dashed blue lines are the cosmological horizons ($r=1$ or $UV=0$), orange curves represent spacelike extremal curves  corresponding to the timelike subregion, and bold green curve represent the associated timelike extremal curve ($U=V$).}
    \label{fig:Timelike_complexity1}
\end{figure}   
\end{center}
In the present case, the boundary (stretched horizon) is two-dimensional, a spatial circle times the time direction, $S^1\times \mathbb{R}$. To study timelike complexity, we consider the timelike boundary subregion shown in Fig.~(\ref{fig:Timelike_complexity1}), defined by the interval $-T \leq t \leq T$ for fixed $\phi=\phi_0$. We first determine the corresponding codimension-two spacelike and timelike extremal surfaces. Using the spherical symmetry of the spacetime, the extremal surface can be parametrized as $U=U(V)$. The induced metric on this codimension-two surface is given by
\begin{equation}
ds^{2}_{\text{ind}} = \frac{-4\,U'(V)}{\bigl(1 - U(V)\,V\bigr)^2}\, dV^2\,.
\end{equation}
From this induced metric, we obtain the length functional to be extremized,
\begin{equation}
\mathcal{L} = \int dV \frac{2\sqrt{-U'(V)}}{1 - U(V)\,V}\,.\label{area}
\end{equation}
The Euler--Lagrange equation representing the extremal surface (RT curve) is
\begin{equation}
\bigl(1 - V U(V)\bigr)\,U''(V)
+ 2V \bigl(U'(V)\bigr)^2
- 2U(V)\,U'(V)
= 0\,.
\end{equation}
The general solution to the above differential equation
can be written as
\begin{equation}\label{eq:ext_sol}
U(V)=\frac{a V-b^2 V-b}{1+b V}
=\frac{aV}{1+b V}-b
\end{equation}
where $a,b$ are constants of integration. Differentiating, we obtain
\begin{equation}
U'(V)=\frac{a}{(1+b V)^2}\,.
\end{equation}
The sign of the constant \(a\) determines the nature of the extremal surface: for \(a>0\), the surface is timelike, while for \(a<0\), it is spacelike. We work out the two cases separately in the following subsections.

\subsubsection{The spacelike extremal curve}
First, we consider the case of a spacelike extremal curve, which we denote by
\begin{equation}
U(V)=\frac{-aV}{1+b V}-b\,,\quad \text{with}\quad a>0\,. \label{spacelike}
\end{equation}
From the symmetry of the diagram Fig.~(\ref{fig:Timelike_complexity1}), the turning point of the extremal curve lies on the line \(U=V\), corresponding to the boundary time \(t=0\) where the Schwarzschild coordinate \(r\) is maximized. We denote this turning point by \((U^{*},V^{*})\). The Schwarzschild coordinate is encoded implicitly through the relation
$UV=e^{-2\tilde r(r)}$,
where \(\tilde r(r)\) denotes the tortoise coordinate. Maximizing $r$, is therefore equivalent to maximizing (extremizing) the quantity \(UV\)
\begin{equation}
\frac{d}{dV}\bigl(U(V)\,V\bigr) = 0\,.
\end{equation}
Considering the spacelike RT curves to begin and end on the symmetric points on the two stretched horizons, say $(r,T)$ and $(r,-T)$ \cite{Jorstad:2022mls}, at the turning point where $r$ (equivalently $UV$) attains a maximum in region $II$ of Fig.~(\ref{P1}), the coordinates must satisfy \(U = V\). This condition implies the relation between the constants,
\begin{equation}
a = 1 - b^2.
\end{equation}
As for spacelike surface $a>0$ this restrict $-1<b<1$. This condition leads to
\begin{equation}\label{eq: extremalpoints}
U^* = \frac{\sqrt{1-b^2}-1}{b}\,, 
\qquad
V^* =\frac{\sqrt{1-b^2}-1}{b}\,,
\qquad \text{with } -1<b < 0\,.
\end{equation}
As the numerator is negative, one must have $b<0$ in order for both $U^{*}$ and $V^{*}$ to be positive, as is the case in region $II$ of Fig.~(\ref{P1}). Also reality forces $b^2<1$, so one must have $-1<b<0$. Substituting this relation back into the solution, the profile of the surface simplifies to
\begin{equation}
U(V) = \frac{-(b+V)}{1+bV}\,.
\end{equation}
Equivalently, this expression can be inverted to yield
\begin{equation}
V(U) = \frac{-(b+U)}{1+bU}\,.
\end{equation}
Imposing the boundary conditions (\ref{boundary}),
we determine the constant \(b\) as
\begin{equation}
b=\frac{\epsilon  \sinh (T)}{(\epsilon -1) \sqrt{\frac{\epsilon }{2-\epsilon }}}\nonumber \approx -\sqrt{2\epsilon}\sinh{T}\,.
\end{equation}
The length of the spacelike extremal curve is given by (\ref{area}), 
\begin{align}
    A^{(s)}=& \int_{-U^0}^{U^*} dU \, \frac{2\sqrt{-V'(U)}}{1 - U\,V(U)}\,,\\
    =&2 \tan ^{-1}\left(\frac{\left(\sqrt{1-b^2}+1\right) \sqrt{\frac{\epsilon }{2-\epsilon}}-b e^T}{\left(\sqrt{1-b^2}+1\right) e^T-b \sqrt{\frac{\epsilon }{2-\epsilon }}}\right)\,.
\end{align}

\subsubsection{The timelike extremal curve}
Now we consider the case of a timelike extremal curve. We denote the timelike extremal curve by
\begin{equation}
U(V)=\frac{cV}{1+d V}-d,\quad\text{with $c>0$} \label{timelike geodesic 3}
\end{equation}
where $c,d$ are a pair of integration constants (distinct from $a,b$ in (\ref{spacelike})). To determine the constants $c,d$, i.e. to determine the timelike extremal curve associated with the timelike subregion, we call upon the spacelike extremal curves anchored to the extremities of the timelike subregion on the stretched horizon, with turning points $(U^*,V^*)$ and $(-U^*,-V^*)$, refer to \eqref{eq: extremalpoints},  (owing to the symmetry of the Fig.~(\ref{fig:Timelike_complexity1})). The timelike extremal curve is obtained by connecting these two turning points with a timelike geodesic \eqref{timelike geodesic 3}. Imposing these two boundary conditions uniquely fixes the constants to be
$c=1$ and $d=0$.
Consequently, the timelike curve reduces to the simple form

\begin{equation}
U(V) = V \quad \text{or} \quad V(U) = U\,.
\end{equation}
The length of the timelike extremal curve is given by, 
\begin{align}
    A^{(t)}=& \int_{-U^*}^{U^*} dU \, \frac{2\sqrt{V'(U)}}{1 - U\,V(U)}\,,\\
=&\int_{-U^*}^{U^*} dU \, \frac{2}{1 - U^2}\,,\\
=&4 \tanh ^{-1}\left(\frac{\sqrt{1-b^2}-1}{b}\right)
\end{align}
where $U^*$ and $V^*$ are defined in (\ref{eq: extremalpoints}).\\

Finally, the total length of the extremal curve is given by \cite{Doi:2023zaf},
\begin{align}
    A= &2A^{(s)}+i A^{(t)}\nonumber,\\
    =&4 \tan ^{-1}\left(\frac{\left(\sqrt{1-b^2}+1\right) \sqrt{\frac{\epsilon }{2-\epsilon}}-b e^T}{\left(\sqrt{1-b^2}+1\right) e^T-b \sqrt{\frac{\epsilon }{2-\epsilon }}}\right)+4i \tanh ^{-1}\left(\frac{\sqrt{1-b^2}-1}{b}\right)\,.
\end{align}
Note that the timelike part of the segment contributes to the length with an imaginary factor of $i$. From this, the timelike entanglement entropy follows \cite{Doi:2023zaf, Doi:2022iyj},
\begin{align}
    S=\frac{A}{4 G_{N}}=&\frac{1}{G_{N}} \tan ^{-1}\left(\frac{\sqrt{2} \left(\sqrt{1-b^2}+1\right) \sqrt{\epsilon }-2 b e^T}{2 \left(\sqrt{1-b^2}+1\right) e^T-\sqrt{2} b \sqrt{\epsilon }}\right)+ \frac{i}{G_{N}} \tanh ^{-1}\left(\frac{ \sqrt{1-b^2}-1}{b}\right)\,,\nonumber\\
    \simeq&-\frac{1}{G_{N}}\tan ^{-1}\left(\frac{b}{\sqrt{1-b^2}+1}\right)+ \frac{i}{G_{N}} \tanh ^{-1}\left(\frac{\sqrt{1-b^2}-1}{b}\right)+\mathcal{O}(\sqrt{\epsilon})\,.
\end{align}
The first line of the above equation gives the exact expression for the timelike entanglement entropy, while the second line corresponds to its expansion in terms of  $\epsilon\to0$ limit at finite $b$. As $\epsilon\to0$, and $b\to0$ one finds that the entanglement entropy also approaches zero. Evidently, as $\epsilon\to0$, both $U^0$ and $V^0$ approach zero, indicating that the two boundary points on the stretched horizon merge at the bifurcation circle ($S^1$) in the Penrose diagram shown in Fig.~(\ref{fig:Timelike_complexity1}). Therefore, to obtain a finite entanglement entropy, $b=-\sqrt{2\epsilon}\,\sinh T$ must remain finite. This can happen if $T$ becomes large and simultaneously $\epsilon \rightarrow 0\,.$ Finally, the lower bound $-1<b$ leads to the following upper bound on the time,
$T_{\textrm{max}}$
\begin{equation}\label{eq: IR_cutoff}
    T_{\textrm{max}}=  \frac{1}{2}\ln\left(\frac{2-\epsilon}{\epsilon}\right)\,.
\end{equation}
Now, we draw attention to a couple of key features of the timelike entanglement entropy expression obtained above. 
\begin{itemize}

 \item As the stretched horizon approaches the cosmological horizon ($\epsilon\to0$), the timelike entanglement entropy remains finite, with neither the real nor the imaginary part exhibiting any divergence. We also find that, as $\epsilon \to 0$, both the real and imaginary parts remain positive.

 \item The complex nature of the timelike entanglement entropy is consistent with previous studies of timelike entanglement entropy and holographic pseudo-entropy \cite{Doi:2023zaf,Doi:2022iyj}. Unlike holographic pseudo-entropy \cite{Doi:2023zaf}, which typically exhibits a divergence as the cutoff goes to zero, the timelike entanglement entropy in our setup remains finite in the limit $\epsilon \to 0$. 
    
\end{itemize}

\subsection{Timelike Volume Complexity}
In this section, we compute the timelike volume complexity for the pure dS geometry utilizing the proposal of \cite{Alishahiha:2025xml}, which was mainly investigated for the asymptotically AdS spacetime. It is given as,
\begin{align}
    C_{s}={\frac{1}{G_N}}\int d\mathcal{V}_{\Sigma}
\end{align}
where $\mathcal{V}_{\Sigma}$ is the volume of the bulk region, which is bounded by spacelike and timelike extremal curves computed in the previous section.  Here we note that in the case of AdS, the spacelike extremal curves (emanating from the future and past extremities of the boundary timelike subregion), and associated timelike extremal curve, intersect the two spacelike extremal surfaces at future and past infinity, respectively \cite{Alishahiha:2025xml}. In our de Sitter construction, the spacelike extremal curve from the future (past) extremity of the timelike subregion does not reach future (past) infinity, and instead there is a turning point which lies to the future (past) of all other points as depicted in the Fig.~(\ref{fig:Timelike_complexity1}). So, in the dS case, the natural timelike extremal curve necessary for extending the timelike subregion complexity proposal of \cite{Alishahiha:2025xml} to the de Sitter case should be the one that joins the two extremal (maxima and minima in time) or turning points of the two spacelike extremal curves. This region can be divided into three region as shown in Fig.~(\ref{fig:Timelike_complexity1})\,.
\begin{align}
    C_{s}=&\frac{1}{G_{N}}\int \frac{2dU dV}{(1-UV)^{2}}\,,\nonumber\\
    =&\frac{2}{G_{N}}\int_{-\sqrt{\frac{\epsilon}{2-\epsilon}}}^{-U^{0}}dU\int_{-U}^{-\frac{\epsilon}{(2-\epsilon)U}}dV\frac{2}{(1-UV)^{2}}+\frac{2}{G_{N}}\int_{-U^{0}}^{0} dU\int_{-U}^{V_{s}(U)} dV \frac{2}{(1-UV)^{2}}\nonumber\\&+\frac{2}{G_{N}}\int^{U^{*}}_{0} dU\int_{U}^{V_{s}(U)} dV \frac{2}{(1-UV)^{2}}\,,\\
    =&\frac{2T \epsilon}{G_{N}} +\frac{2}{G_{N}}\ln \left(\frac{-2 b e^{-T} \sqrt{2-\epsilon } \sqrt{\epsilon }+e^{-2 T} \epsilon -\epsilon +2}{2 \sqrt{1-b^2}}\right)\,,\nonumber\\
    =&\frac{2T \epsilon}{G_{N}} +\frac{2}{G_{N}}\ln \left(\frac{ 2-\epsilon-\epsilon\,e^{-2 T}}{2 \sqrt{(1-\epsilon )^2-(2-\epsilon) \epsilon\,  \sinh ^2(T)}}\right)\,.
\end{align}
In the second line, the factor of 2 has been included on account of the symmetry of de Sitter geometry across $t=0$ ($U\leftrightarrow V$). Using the holographic dual interpretation of Newton's constant $G_{N}$ as the Brown-Henneaux central charge of a putative boundary theory  $c=\frac{3}{2 \,G_N}$, the timelike complexity turns out to be proportional to the central charge $c$.
\begin{align}
    C_{s}=\frac{4c}{3}T\epsilon+\frac{4c}{3}\ln \left(\frac{ 2-\epsilon-\epsilon\,e^{-2 T}}{2 \sqrt{(1-\epsilon )^2-(2-\epsilon) \epsilon\,  \sinh ^2(T)}}\right)\,.
\end{align}
Here we list the salient features of the timelike subregion complexity expression obtained above.
\begin{itemize}
\item From the expression of complexity in the second line above, it can be seen that for fixed $b$ in a small $\epsilon$ limit, the volume contribution from region I and II vanish as $\mathcal{\epsilon}$ and $\mathcal{\sqrt{\epsilon}}$ respectively.

    \item In the limiting case of $\epsilon\rightarrow0$ and finite $b$ the complexity is 
\begin{align}
    C_{s}\approx-\frac{4c}{3}\frac{1}{2} \ln \left(1-b^2\right)-\frac{4c}{3}\sqrt{2} b e^{-T}\sqrt{\epsilon }+\epsilon \frac{4c}{3} \left(\frac{e^{-2 T}}{2}-b^{2}e^{-2T}+T-\frac{1}{2}\right)+O\left(\epsilon ^{3/2}\right)\,.
\end{align}
\item Notice that as $b^{2}<1$, the leading contribution to timelike complexity is positive and real.
\item This expression does not exhibit any UV divergences; the timelike complexity is finite, real, and positive. This stands in stark contrast to the AdS case \cite{Alishahiha:2025xml}.

\item To study the early-time behavior of the complexity, we rewrite it in terms of the maximal time $T_{\max}$ as
\begin{align}
    C_{s}&=\frac{4c}{3}T\epsilon+\frac{4c}{3}\ln\Bigg(\frac{1-e^{-2(T_{\textrm{max}}+T)}}{\sqrt{1-\left(e^{2(T-T_{\textrm{max}})}+e^{-2(T_{\textrm{max}}+T)}-2e^{-T_{\textrm{max}}}\right)}}\Bigg),\\
    &\simeq\frac{2c}{3}e^{2(T-T_{\textrm{max}})},\,\,\quad e^{2(T-T_{\textrm{max}})}\ll1\,.
\end{align}
The above expression reveals that the subregion timelike volume complexity grows exponentially for $T\ll T_{\textrm{max}}$, i.e. early time regime with a growth rate $2$ in units of the dS radius.

\item In the late-time regime, the subregion timelike volume complexity displays a hyperfast growth, it diverges logarithmically as $T\rightarrow T_{\rm max}$,
\begin{equation}
   C^{\rm late}_{s}\sim \frac{2c}{3}\ln \left[ \left(T_{\rm max}-T\right)^{-1} \right].
\end{equation}
This logarithmic divergence is actually a dimensional accident, for higher dimensions the degree of divergence is $(d-2)$ as will be demonstrated in the Section~(\ref{sec:dS_d}).

\item We emphasize here once again that the constant $b$ is negative, and the requirement of the turning point coordinates $U^{*}$ and $V^{*}$ to be real puts a constraint on $b$ that $b^{2}<1$ as in \eqref{eq: extremalpoints}. As the constant $b=\frac{\epsilon  \sinh (T)}{(\epsilon -1) \sqrt{\frac{\epsilon }{2-\epsilon}}}$ this bound translates into an upper bound on the time $T$ as given in \eqref{eq: IR_cutoff}. 
\end{itemize}

\section{Higher dimensional de Sitter}\label{sec:dS_d}
In this section, we generalize our treatment to arbitrary dimensions, i.e., we consider a general $(d+1)$-dimensional de Sitter spacetime. In static coordinates, the metric reads,
\begin{align}
ds^2 = -(1 - r^2)\,dt^2 + \frac{dr^2}{1 - r^2} + r^2 d\Omega_{d-1}^2\,.
\end{align}
while in Kruskal coordinates the metric is given by,
\begin{align}
    ds^2 = \frac{-4}{(1-UV)^2} \, dU\, dV 
+ \frac{(1+UV)^2}{(1-UV)^2} \, d\Omega_{d-1}^2\,.
\end{align}
To study timelike complexity \emph{i.e.} complexity associated with a boundary subregion extended in the time direction, we first consider the task of finding codimension two spacelike and timelike extremal surfaces corresponding to the timelike boundary subregion $-T\leq t\leq T$ at a constant $\phi=\phi_{0}$. Owing to spherical symmetry, the extremal surface will be of the form $V=V(U),\,\phi=\phi_0$. The induced metric on this codimension two surface is given by
\begin{align}
    ds^2 = \frac{-4}{(1-UV(U))^2} \,V'(U) dU^{2}
+ \frac{(1+UV)^2}{(1-UV)^2} \,\sin^{2}{\phi_{0}} d\Omega_{d-2}^2\,,
\end{align}  
and the area is
\begin{align}\label{eq:area}
    A=&\int d\Omega_{d-1}\int dV\sqrt{g}\,,\nonumber\\
    =&\int d\Omega_{d-1}\int dV\frac{2 (\sin ^{d-2}\phi_{0})\, (U V(U)+1)^{d-2} \sqrt{-V'(U)}}{(1-U V(U)) (1-U V(U))^{d-2}}\,.
\end{align}
The Euler-Lagrange equation resulting from extremizing this functional is,
\begin{align}
    2 (2 d+U V(U)-3) V'(U) \left(V(U)-U V'(U)\right)+\left(U^2 V(U)^2-1\right) V''(U)=0\,.
\end{align}
Notice that for $d=2$, i.e. for $dS_{3}$, we solved the Euler-Lagrange equation analytically and obtained an explicit solution \eqref{eq:ext_sol}. However, for general dimensions, a solution is not tractable analytically. One has two choices here, either to persist with analytical means by making some simplifying assumptions or approximations, or studying some limits, or one can resort to numerical means to obtain an exact solution without approximations. We rely on the latter approach, i.e. solve this equation numerically with appropriate boundary conditions which are well motivated in $d=2$ case. For the spacelike extremal surface we specify one boundary condition at the cosmological horizon as
\begin{align}
V\left(-U^{0}\right)&=V^{0},
\end{align}
$U^{0}$ and $V^{0}$ are defined in \eqref{eq:U0_V0}, and another at the turning point outside the cosmological horizon where $r$ hits a maximum \emph{i.e} the product $UV$ hits an extremum point. Again, as we did for dS$_{3}$ by utilizing the symmetry of geometry in Fig.~(\ref{fig:Timelike_complexity1}), we can say that the turn point in $r$, which we denote by $r_{*}$, should coincide with the boundary time $t=0$. In the Kruskal  coordinates this turn point $\{0,r_{*}\}$ translates into, 
\begin{align}
   V(U^{*})=U^{*}, 
\end{align}
and $U^{*}$ can be determined numerically in terms of $\epsilon$ and $T$ by extremizing the product $UV(U)$
\begin{align}\label{eq:extremalpoint_num}
    \frac{d(UV(U))}{dU}\Bigg|_{U=U^*}=0\,.
\end{align}
Once we have the numerical solution to the Euler-Lagrange equation, we can proceed to determine the area as given in \eqref{eq:area} numerically.
As a consistency check of our numerics a comparison of the numerical estimates with the analytic estimates obtained in Section~(\ref{d=2}) of area of spacelike extremal surface for $d=2$ is provided in Table~(\ref{tab:spacelikearea_num}) \footnote{For the consistency check the numerical values of boundary time $T$ are chosen to be of $\mathcal{O}(1)$ because for a given value of $\epsilon$ there is an upper bound on $T$, which is $\mathcal{O}(1)$ for $\epsilon\sim 10^{-7}$. Also, for small values of $T$, the domain of numerical solution of $V(U)$ shrinks, which leads to precision errors.}.
\begin{table}[]
    \centering
    \begin{tabular}{|c|c|c|}
        \hline
$T$& $A^{s}$ & $A_{s}^{(\text{num})}$\\
\hline
1 & 6.90$\times10^{-4}$ & 6.80$\times10^{-4}$\\ 
\hline
2& 1.68$\times10^{-3}$ & 1.68$\times10^{-3}$\\
\hline
3& 4.50$\times10^{-3}$&  4.50$\times10^{-3}$\\
\hline
4& 1.22$\times10^{-2}$& 1.22$\times10^{-2}$\\ 
\hline
5& 3.32$\times10^{-2}$&  3.32$\times10^{-2}$\\
\hline
6& 9.00$\times10^{-2}$& 9.00$\times10^{-2}$\\
\hline
7& 0.25&0.25 \\
\hline

    \end{tabular}
    \caption{A comparison of numerical and analytic estimates for the area of a spacelike extremal surface for $2+1$-dimensional de Sitter spacetime with fixed $\epsilon=10^{-7}$ and $\phi_{0}=\frac{\pi}{2}\,.$}
    \label{tab:spacelikearea_num}
\end{table}

\begin{table}[htb!]
    \centering
    \begin{tabular}{|c|c|c|}
   \hline
   $T$ &  $A^{t}$ & $A_{t}^{(\text{num})}$\\
   \hline
 1 & 1.05$\times10^{-3}$ & 1.26$\times10^{-3}$ \\
 \hline
 2 & 3.24$\times10^{-3}$ & 3.26$\times10^{-3}$ \\
 \hline
 3 & 8.96$\times10^{-3}$ & 8.92$\times10^{-3}$ \\
 \hline
 4 & 2.44$\times10^{-2}$ & 2.44$\times10^{-2}$ \\
 \hline
 5 & 6.64$\times10^{-2}$ & 6.64$\times10^{-2}$ \\
 \hline
 6 & 0.18091 & 0.18091 \\
 \hline
 7 & 0.500629 & 0.500629 \\
 \hline
    \end{tabular}
    \caption{A comparison of numerical and analytic estimates for the area of a timelike extremal surface for $2+1$-dimensional de Sitter spacetime with fixed $\epsilon=10^{-7}$ and $\phi_{0}=\frac{\pi}{2}\,.$}
    \label{tab:timelikearea_num}
\end{table}

For the timelike extremal surface we solve the Euler-Lagrange equation numerically using the following two boundary conditions,
\begin{align}
    V(U^{*})=U^{*},\quad V(-U^{*})=-U^{*}
\end{align}
This obviously results into a ``straight line" solution,where $U^{*}$ is given by, \eqref{eq:extremalpoint_num}. The area of the extremal timelike surface is then determined numerically for a $(d+1)$ dimensional de Sitter spacetime. A comparison of numerical and analytic results for $2+1$-dimesional de Sitter spacetime is presented in Table~(\ref{tab:timelikearea_num}). 

Now we possess all the necessary ingredients to determine complexity for the timelike subregion $-T\leq t \leq T, \phi=\phi_0$. Just as was done for $2+1$-dimensional de Sitter spacetime, the timelike subregion complexity is determined by the spacetime volume confined between the spacelike and timelike extremal surface computed here (up to a factor of $G_N$),
\begin{align}
    C_{s}=&{\frac{1}{G_N}}\int d\mathcal{V}_{\Sigma}\,,\nonumber\\
    =&\frac{2 \sin ^{d-2}\phi_{0}\Omega_{d-2}}{G_{N}}\int dU \int dV\frac{ (U V+1)^{d-2}}{(1-U V)^2 (1-U V)^{d-2}}\,.
\end{align}
The volume can be calculated by dividing the spacetime into three regions as shown in Fig.~(\ref{fig:Timelike_complexity1}). 
\begin{align}
       C_{s} =&\frac{2}{G_{N}}\int_{-\sqrt{\frac{\epsilon}{2-\epsilon}}}^{-U^{0}}dU\int_{-U}^{-\frac{\epsilon}{(2-\epsilon)U}}dV~\mathcal{V}_{\Sigma}+\frac{2}{G_{N}}\int_{-U^{0}}^{0} dU\int_{-U}^{V_{s}(U)} dV ~\mathcal{V}_{\Sigma}\nonumber\\&+\frac{2}{G_{N}}\int^{U^{*}}_{0} dU\int_{U}^{V_{s}(U)} dV~ \mathcal{V}_{\Sigma}\,.
\end{align}
Here we substitute the numerical solution for the spacelike extremal surface $V_{s}(U)$ and $U^{*}$, and subsequently perform the spacetime volume integrals numerically. This determines the timelike complexity of $d+1$-dimensional de Sitter spacetime numerically. We have presented the comparative results for 2+1-dimensional de Sitter spacetime in Table~(\ref{tab:timelikecomplexity_num}). Again, we are bound to $\mathcal{O}(1)$ numerical values of boundary subregion time $T$ due to the validity of the numerical solution of the Euler-Lagrange equation. 
\begin{table}[htb!]
    \centering
    \begin{tabular}{|c|c|c|}
    \hline
    $T$ & $C_s$ & $C_{s}^{\text{num}}$\\
    \hline
 1 & 6.36$\times10^{-7}$ & 5.63$\times 10^{-7}$ \\
 \hline
 2 & 3.14$\times10^{-6}$ & 3.12$\times10^{-6}$ \\
 \hline
 3 & 2.07$\times10^{-5}$ & 2.08$\times10^{-5}$ \\
 \hline
 4 & 1.50$\times10^{-4}$ & 1.50$\times10^{-4}$ \\
 \hline
 5 & 1.10$\times10^{-3}$ & 1.10$\times10^{-3}$ \\
 \hline
 6 & 8.17$\times10^{-3}$ & 8.17$\times10^{-3}$ \\
 \hline
 7 & 6.2$\times10^{-2}$ & 6.2$\times10^{-2}$ \\
\hline
    \end{tabular}
    \caption{A comparison of numerical and analytic results for timelike complexity for  $2+1$-dimensional de Sitter spacetime with $\epsilon=10^{-7}$, $\phi_{0}=\frac{\pi}{2}$ and $G_{N}=1\,.$ }
    \label{tab:timelikecomplexity_num}
\end{table}

\begin{figure}
    \centering
    \includegraphics[width=0.6\linewidth]{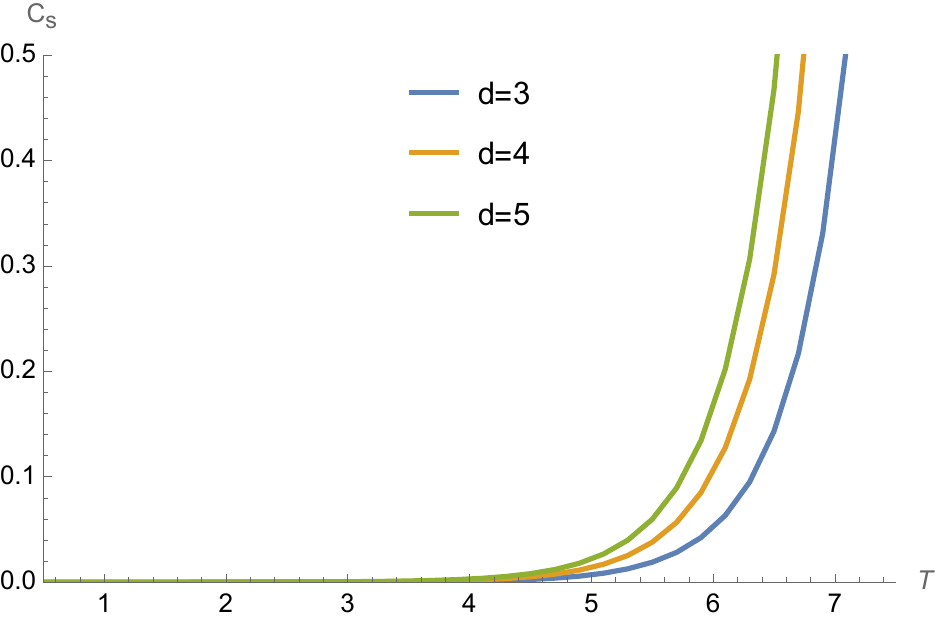}
    \caption{A numerical plot for timelike complexity $C_{s}$ with boundary subregion extension (time) $T$ for $d=3,4,5$ dimensional de Sitter spacetime with $\epsilon=10^{-7}$, $\phi_{0}=\frac{\pi}{2}$ and $G_{N}=1\,.$}
    \label{fig:complexity_num}
\end{figure}
As the numerical result is valid for all higher-dimensional de Sitter spacetimes, we study the dependence of timelike complexity on the boundary subregion time $T$ for various dimensions. This is shown in the numerical plots in Fig.~(\ref{fig:complexity_num}). 
\\

\begin{figure}
    \centering
    \includegraphics[width=0.6\linewidth]{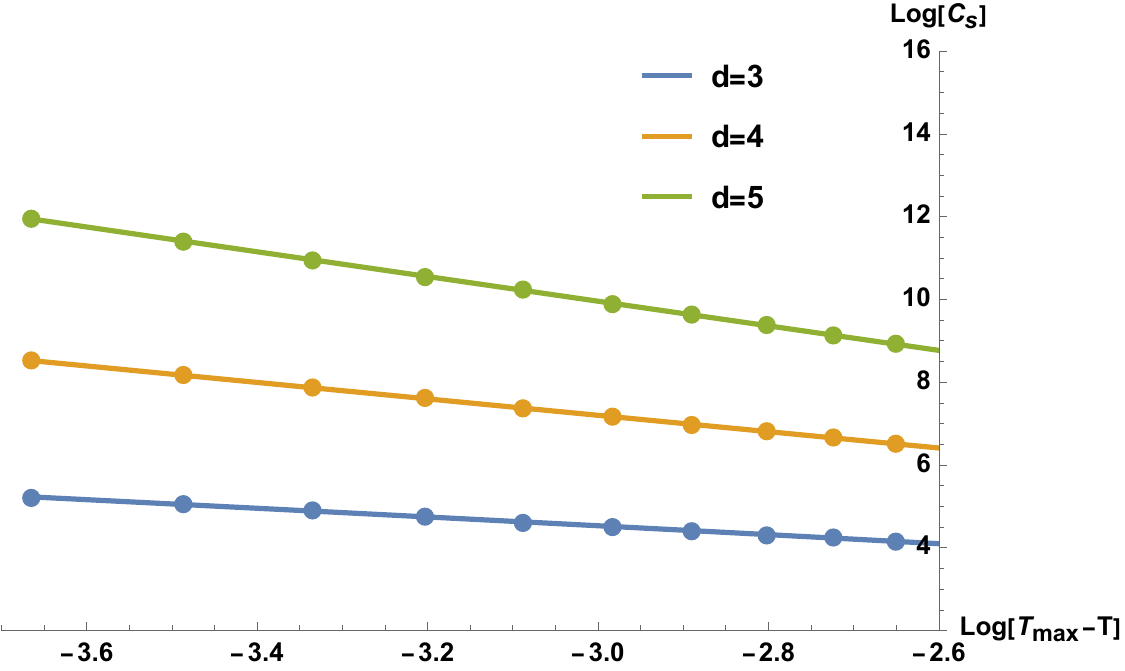}
    \caption{The numerical plot for $\ln{C_{s}}$ with $\ln{(T_{\text{max}}-T)}$ for $d=3,4,5$ dimensional de Sitter spacetime with $\epsilon=10^{-7}$, $\phi_{0}=\frac{\pi}{2}$ and $G_{N}=1$. The plot represents a power law dependence on time $T_{\text{max}}-T$ with a dimension-dependent exponent $(d-2)$. This late-time feature is the same as was observed for spacelike volume complexity and spacelike subregion complexity.}
    \label{fig:logCvslogT_latetime}
\end{figure}

For the late time regime, it can be seen from the numerical plots shown in Fig.~(\ref{fig:complexity_num}) that, similar to its spacelike counterpart, the timelike complexity also exhibits indications of an \textit{hyperfast growth}. By a focused numerical analysis on the late time regime, we have found that the boundary subregion time attains a maximum value that is independent of the spacetime dimension $d$, and is given by,
\begin{align}
    T_{\text{max}}=\frac{1}{2}\ln\left(\frac{2-\epsilon}{\epsilon}\right)\
\end{align}
and as it can be seen clearly from the numerical plots shown in Fig.~(\ref{fig:logCvslogT_latetime}) in the late time regime, the timelike complexity behaves as,
\begin{align}
    C_{s}\sim \frac{1}{(T_{\text{max}}-T)^{d-2}}\,.
\end{align}

\begin{figure}
    \centering
    \includegraphics[width=0.6\linewidth]{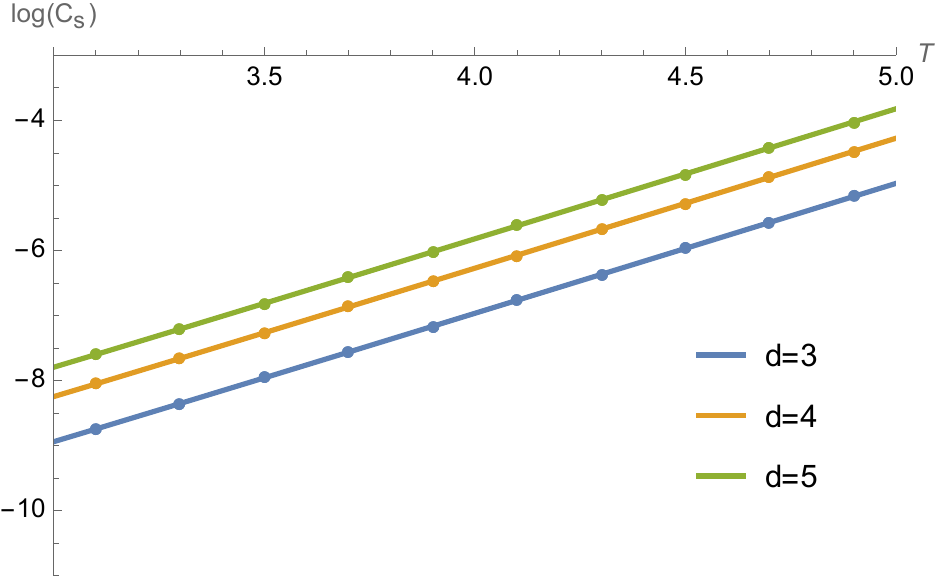}
    \caption{The numerical plot for $\ln{C_{s}}$ with boundary subregion extension (time) $T$ for $d=3,4,5$ dimensional de Sitter spacetime with $\epsilon=10^{-7}$, $\phi_{0}=\frac{\pi}{2}$ and $G_{N}=1$. The plot represents an exponential dependence on time $T$ with a dimension-independent exponent $\sim 2$.}
    \label{fig:logCvsT_earlytime}
\end{figure}
 We have also performed a numeric analysis focusing on early time behavior of timelike complexity, which exhibits an \textit{exponential growth} as can be seen from the numerical plot presented in Fig.~(\ref{fig:logCvsT_earlytime}). From these numerical plots of $\ln(C_{s})$ with boundary time $T$ for different spacetime dimensions, we can read off the exponents which turn out to be independent of the spacetime dimension $d$ and have a fixed numerical value of two. 

\begin{equation}
    C_S \sim e^{2T}.
\end{equation}

\section{Schwarzschild de Sitter black holes}\label{sec:SdS_BH}
The $(d+1)$-dimensional Schwarzschild de Sitter (SdS) black hole metric in static coordinates \cite{Spradlin:2001pw}, and in units where dS radius is set to unity, reads,
\begin{equation}
    ds^2=-\left(1-\frac{\mu}{r^{d-2}}-r^2\right) dt^2+\left(1-\frac{\mu}{r^{d-2}}-r^2\right)^{-1} dr^2+r^2\,d\Omega^2_{d-1}\,,
\end{equation}
where $\mu=16 \pi G_N M/V_{S^{d-1}}$ is a parameter proportional to the black hole mass $M$, the $(d+1)$-dimensional Newton's constant up to dimension-dependent geometric factors. The case of $d=2$, i.e., three dimensions, is a bit exceptional since there are no black holes in Einstein theory with a positive cosmological constant. The metric in this case reads,
\begin{equation*}
    ds^2=-\left(1-8\pi G_N M-r^2\right) dt^2+\left(1-8\pi G_N M-r^2\right)^{-1}dr^2+r^2\,d\phi^2.
\end{equation*}
Redefining $\sqrt{1-8 \pi G_N M}=l_c$, the metric reads,
\begin{equation}
    ds^2=-\left(l_c^2-r^2\right) dt^2+\frac{dr^2}{l_c^2-r^2}+r^2\,d\phi^2. \label{SdS_3}
\end{equation}
Note that $0<l_c<1$. In this form of the metric, it is evident that this is not a black hole but a conical defect spacetime with the defect located at $r=0$ and the deficit angle being $\Delta\varphi = 2\pi (1-l_c)$. The timelike subregion volume complexity for this exceptional case is treated separately in Section~(\ref{con def}).
\\

\begin{figure}[h]
    \centering
    \includegraphics[width=0.9\linewidth]{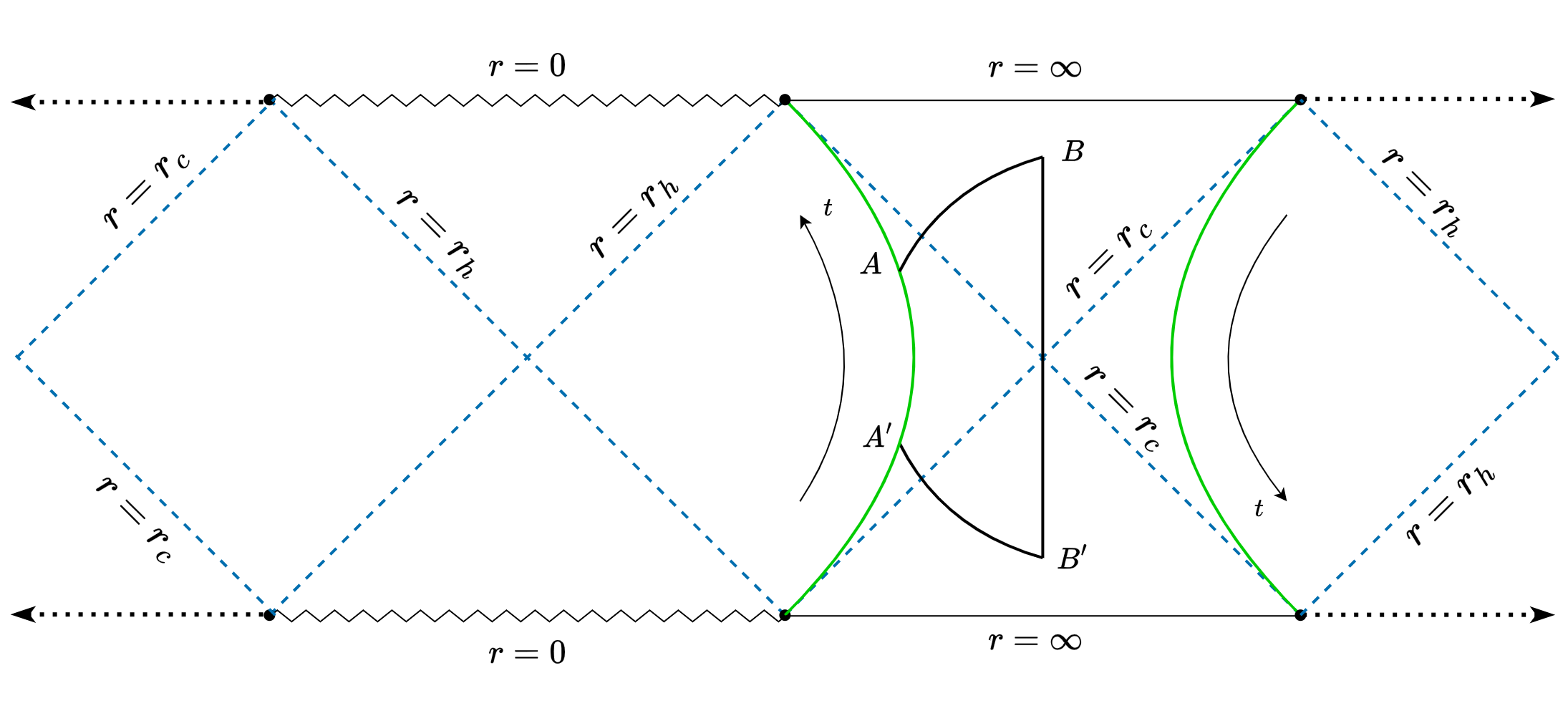}
    \caption{Penrose diagram of the infinitely extended Schwarzschild--de Sitter spacetime. The blue dashed lines correspond to the black hole horizon ($r=r_h$) and the cosmological horizon ($r=r_c$), while the green curves denote the stretched horizons for case 1, located at $r=r_c-\epsilon$ and $r=r_c+\epsilon$, respectively. The curvature singularity is at $r=0$, and the segments labeled ($r=\infty$) represents future (past) infinities. The curves $AB$ and $A'B'$ are spacelike extremal surfaces, whereas $BB'$ is a timelike extremal surface. Together, they form the bulk extremal hypersurface used in the computation of the timelike subregion complexity.The arrow shows the direction of increasing Schwarzschild time $t$.}
    \label{fig:BH1}
\end{figure}

In this section we will study timelike complexity of $(d+1)$-dimensional Schwarzschild de Sitter (SdS) black hole for $d\geq3$. For these $(d+1)$-Schwarzschild de Sitter black hole there are two horizons: cosmological ($r_{c}$) and black hole ($r_{h}$) provided the black hole mass parameter $\mu$ is less than or equal to $\mu_{\text{max}}=\frac{1}{3\sqrt{3}}\approx0.19$\footnote{This bound on mass of black hole is specific to the case when the dS radius $L=1$. Later, for numerical convenience, we consider different values of dS radius, $L\neq1$. In that case, the black hole mass bound translates to a constraint on the ratio $\frac{\mu}{L}\leq \frac{1}{3\sqrt{3}}\,.$}. The geometry is infinitely extended in the spacelike direction as it is represented in the Penrose diagram shown in Fig.~(\ref{fig:BH1}).
 When $\mu\leq\mu_{\text{max}}$ we consider the cosmological horizon to be larger (to the exterior of) the black hole horizon, and as the mass parameter reaches its maximum value  $\mu=\mu_{\text{max}}$, both the horizons coincide. This limiting case is the well-known Nariai limit \cite{Spradlin:2001pw, Galante:2023uyf, Anninos:2012qw, Bousso:2002fq}.
\par
As the Schwarzschild de Sitter black hole is infinitely extended, there is more freedom in placing the stretched horizon \emph{i.e}, where the boundary dual field theory lives, and in general, there are various possible configurations \cite{Gibbons:1977mu,Aguilar-Gutierrez:2024rka}. In this paper, we will consider two different configurations. In the first configuration we consider, the stretched horizon lies near the cosmological horizon as shown in Fig.~(\ref{fig:BH1}), and in the second configuration we consider the stretched horizon lies near the black hole horizon as shown in Fig.~(\ref{fig:BH2}). If $r_{\textrm{st}}$ is the Schwarzschild radial coordinate corresponding to the stretched horizon, then for the two cases we consider, one has,
\begin{align}
    r_{\text{st}}=&\rho r_{c}+(1-\rho)r_{h},\quad \text{Case 1}:\rho=1-\epsilon\,,\\
    r_{\text{st}}=&\rho r_{c}+(1-\rho)r_{h},\,\quad \text{Case 2}:\rho=\epsilon\,.
\end{align}
As each case requires focusing on different regions of the Schwarzschild de Sitter black hole, we present the detailed study of each case separately in the sections below.
\subsection{Case 1} \label{subsec: case1}

When the stretched horizon lies near the cosmological horizon we probe the geometry with outgoing Eddington-Finkelstein (EF) coordinates
\begin{align}
    u=t-\tilde{r}(r)
\end{align}
where the tortoise coordinate is defined as follows,
\begin{align}\label{eq:ref_pt}
    \tilde{r}(r)=&\int_{\infty}^{r} \frac{dy}{f(y)},\quad r>r_{c}\,,\\
    \tilde{r}(r)=&\int_{r_{0}}^{r} \frac{dy}{f(y)},\quad r_{h}<r<r_{c}
\end{align}
where $r_{0}$ is a reference point, we have chosen this to be 
\begin{align}
    \frac{d}{dr}\left(\frac{1}{f(r)}\right)\Bigg|_{r=r_{0}}=0\,.
\end{align}
In outgoing EF coordinate, the metric takes the form
\begin{align}
    ds^{2}=-2du~dr-f(r)du^{2}+r^{2}d\Omega_{d-1}^{2}\,.
\end{align}
To proceed with the timelike complexity computation, we first need to find the codimension two spacelike and timelike extremal surfaces. Therefore, we consider a hypersurface $u(r)$ on a constant $\theta=\theta_{0}$ slice, the induced metric on this is given by,
\begin{align}
      ds^{2}=-(2u'+f(r)u'^{2})dr^{2}+r^{2}\sin^{2}\theta_{0}d\Omega_{d-2}^{2}\,.  
\end{align}
The area of this extremal surface is given by,
\begin{align}
   A= \Omega_{d-2}\sin^{d-2}\theta_{0}\int dr~ r^{d-2}\sqrt{-2u'-f(r)u'^{2}}\,.
\end{align}
Notice that the area functional is independent of $u$, and this leads to a conserved quantity,
\begin{align}
    C_{u}=&\frac{\partial A}{\partial u'}=-\frac{r^{d-2} \left(f(r) u'(r)+1\right)}{\sqrt{-u'(r) \left(f(r) u'(r)+2\right)}}\,.
\end{align}
The presence of this conserved quantity simplifies the situation extremely as it allows us to bypass solving the second-order Euler-Lagrange equation. We can simply determine $u'(r)$ in terms of $C_{u}$, and we get two different solutions,
\begin{align}
    u'(r)=&-\frac{1}{f(r)}\pm\frac{C_{u}}{f(r)\sqrt{C_{u}^2 +r^{2(d-2)} f(r)}}\,.
\end{align}
We focus on a spacelike surface extending outside the cosmological horizon. The conserved charge $C_{u}^{2}>r^{2(d-2)}|f(r)|$ as $f(r)<0$ outside the cosmological horizon. If we impose a boundary condition that there is an extremum (turning point) at $r=r_{*}$
at which $u'(r_{*})\rightarrow-\infty$ then the conserved charge $C_{u}$ get determined in terms of $r_{*}$, namely,
\begin{align}
   C_{u}^{2}=-r_{*}^{2(d-2)}f(r_{*})=r_{*}^{2(d-2)}|f(r_{*})|\,.
\end{align}
Notice that this requirement removes the ambiguity among the two candidate solutions, as only one of them satisfies this condition at $r=r_{*}$ \emph{i.e.}
\begin{align}
    u'(r)=&\frac{1}{|f(r)|}-\frac{C_{u}}{|f(r)|\sqrt{C_{u}^2 -r^{2(d-2)} |f(r)|}},\quad r>r_{c}\,.
\end{align}
Furthermore, similar to the pure dS case, using the insight based on symmetry of the Schwarzschild dS geometry, we can argue that the turning point corresponds to $t=0$ (since the Penrose diagram here is infinitely extended on both sides, this left-right symmetry persists). This determines the boundary value of the extremal surface at the turning point,
\begin{align}
    u(r_{*})=-\tilde{r}(r_{*})\,.
\end{align}
Then the extremal surface is determined as,
\begin{align}
  u(r)=\int_{r_{*}}^{r}dr~u'(r) -\tilde{r}(r_{*})\,.
\end{align}
Still, we are left with one undetermined constant $r_{*}$ which can be determined by supplying another boundary condition at the stretched horizon,
\begin{align}
    T-\tilde{r}(r_{\text{st}})+\tilde{r}(r_{*})=\int_{r_{*}}^{r_{\text{st}}} u'(y) dy\,.
\end{align}
 This determines the turning point coordinate $r_{*}$ in terms of $\epsilon$, and $T$. Here, the integral on the RHS, as well as the complexity integral, are not tractable analytically, even for $d=3$, which is the lowest dimension nontrivial case of de Sitter black holes one has to consider.

\begin{figure}
    \centering
    \includegraphics[width=0.6\linewidth]{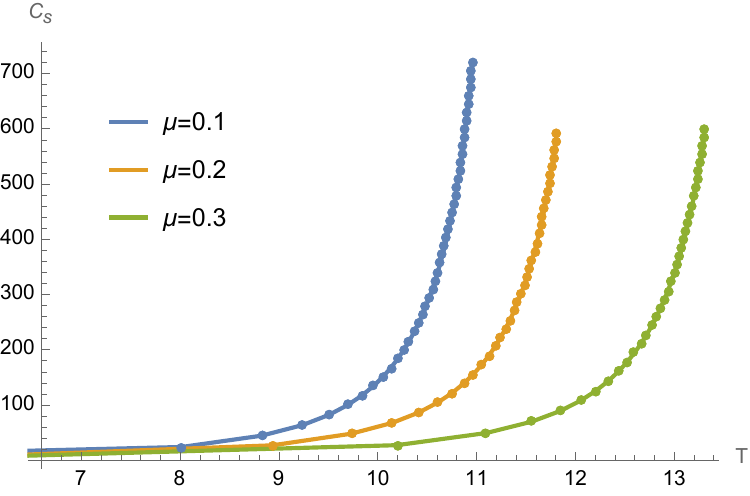}
    \caption{Plot of timelike subregion volume complexity $C_{s}$ for Schwarzschild de Sitter (Case 1) as a function of boundary subregion duration $T$ for $d=3$, $L=3$, and different black hole masses $\mu=0.1,0.2,0.3$ respectively.}
    \label{fig:CvsT_BH1}
\end{figure}

\begin{figure}
    \centering
    \includegraphics[width=0.6\linewidth]{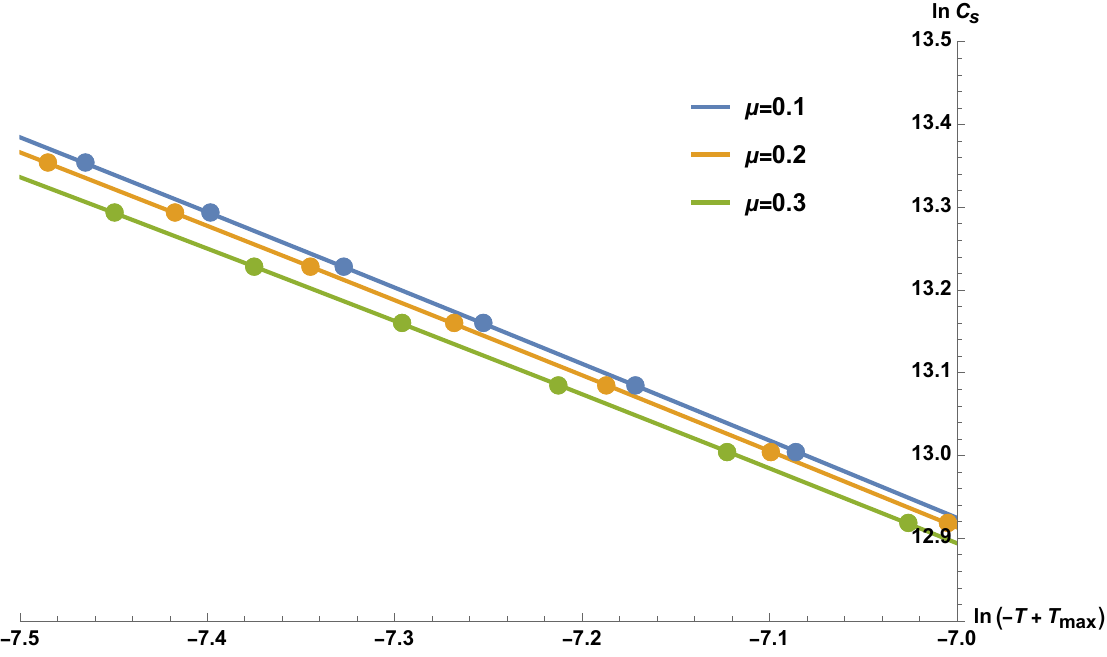}
    \caption{Plot of $\ln{C_{s}}$ with $\ln(T_{\text{max}}-T)$ for $d=3$, $L=3$ and different blackhole masses $\mu=0.1,0.2,0.3$ respectively (Case 1).}
    \label{fig:logCvsT_latetime_BH1}
\end{figure}

\begin{figure}
    \centering
    \includegraphics[width=0.5\linewidth]{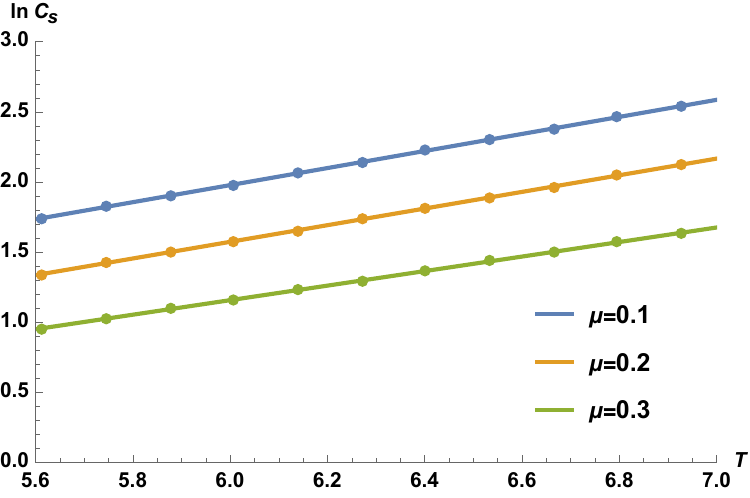}
    \caption{Early boundary time behavior of $\ln{C_{s}}$ for $d=3$ and different blackhole masses $\mu=0.1,0.2,0.3$ respectively (Case 1). Comparing with dS case we observe that even in the presence of blackhole with non-zero mass the early time dependence of timelike complexity is exponential but with an exponent that depends on blackhole mass $\mu$ and AdS radius $L$ and the exponent decreases with increasing numerical values of $\mu$ and $L$.    }
    \label{fig:logCvsT_earlytime_BH1}
\end{figure}

So we proceed numerically to work out the spacelike extremal surface. For the timelike extremal surface which intersects two spacelike surfaces anchored at boundary times $T$ and $-T$ at the pair of turning points $r=r_{*}$, we can argue based on the symmetry of the geometry of SdS (under reflection around $t=0$) that it is given by,
\begin{align}
    u(r)=-\tilde{r}(r)\,.
\end{align}
Now we are equipped with all the ingredients required to compute timelike subregion complexity given by the volume of the region bounded between the spacelike and timelike extremal surfaces\footnote{In the blackhole Case 1 and Case 2 we redefine $C_{s}\sim G_{N}C_{s}$ absorbing the Newton's constant.},
\begin{align}
    C_{s}=2\,\Omega_{d-2}\sin^{d-2}\theta_{0}\left(\int_{r_{\text{st}}}^{r_{c}} dr~ r^{d-2} \int_{\tilde{-r(r)}}^{u_{s}(r)} du+\int_{r_{c}}^{r_{*}} dr~ r^{d-2} \int^{u_{s}(r)}_{u_{t}(r)} du\right)\,.
\end{align}
We perform the integral involved numerically. The numerical behavior of timelike complexity with the boundary subregion time is presented in Fig.~(\ref{fig:CvsT_BH1}) for different choices of black hole masses and for $d=3$. The nature of dependence of timelike complexity over the subregion boundary time is quite similar to the pure de Sitter spacetime. A focused late-time numerical analysis reveals that in this case there is a maximum time $T_{\text{max}}$ as well which is dependent on both the cutoff $\epsilon$ and mass parameter $\mu$, and the late-time timelike complexity exhibits a hyperfast growth as,
\begin{align}
    C_{s}\sim\frac{1}{(T_{\text{max}}-T)^{(d-2)-\#g(\mu)}}\,.
\end{align}
which is exhibited in the plot shown in Fig.~(\ref{fig:logCvsT_latetime_BH1}). The function $g(\mu)$ is a monotonic increasing function of blackhole mass, and it vanishes when $T_{\text{max}}-T$ is of $\mathcal{O}(\epsilon)$. Also from the numerical analysis we have observed that as we increase $\mu$, $T_{\text{max}}$ increases \emph{i.e.} increasing mass of the black hole delays the hyperfast growth. An early time analysis shown in Fig.~(\ref{fig:logCvsT_earlytime_BH1}) shows that the complexity depends exponentially on the boundary subregion time, while the presence of black hole mass is exhibited in the reduction of the exponent compared with the vacuum de Sitter.


\subsection{Case 2}\label{subsec: case2}
\begin{figure}[htbp]
    \centering
    \includegraphics[width=0.9\linewidth]{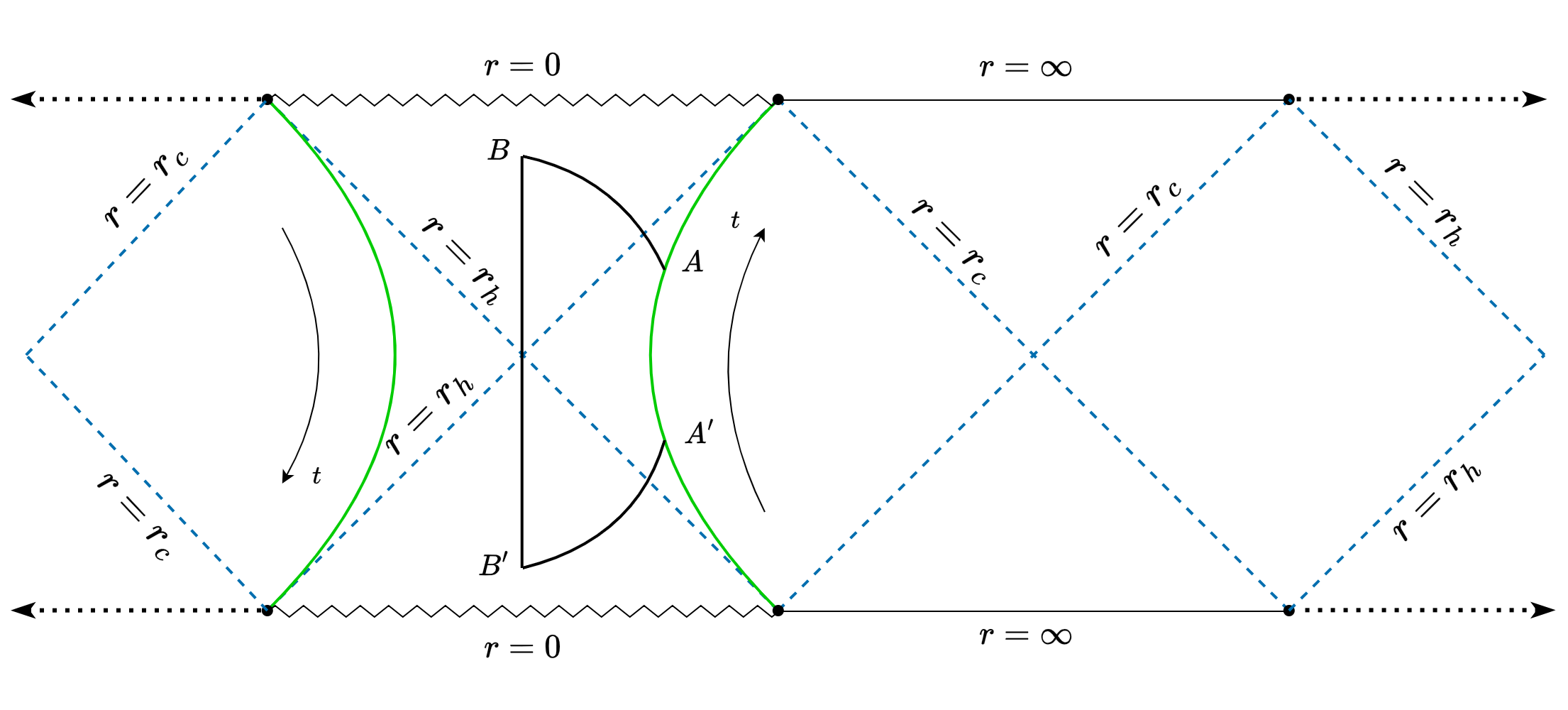}
    \caption{Penrose diagram of the infinitely extended Schwarzschild--de Sitter spacetime, where the stretched horizon, represented by the green curves, is located near the black hole horizon at $r=r_h-\epsilon$ and $r=r_h+\epsilon$. The curves $AB$ and $A'B'$ are spacelike extremal surfaces, whereas $BB'$ is a timelike extremal surface. Together, they form the bulk extremal hypersurface used in the computation of the timelike subregion complexity. The arrow shows the direction of increasing Schwarzschild time $t$.}
    \label{fig:BH2}
\end{figure}

In this case, the boundary CFT lies on a stretched horizon, which we consider near the black hole horizon $r_{h}$, and the spacelike extremal surface extends to the black hole interior, as shown in Fig.~(\ref{fig:BH2}). To probe this region, we consider the ingoing Eddington–Finkelstein coordinates
\begin{align}
    v=t+\tilde{r}(r)
\end{align}
where,
\begin{align}
    \tilde{r}(r)=\int_{r_{0}}^{r} \frac{dy}{f(y)},\quad r>r_{h}\,,\\
    \tilde{r}(r)=\int_{0}^{r} \frac{dy}{f(y)},\quad r<r_{h}
\end{align}
where, $r_{0}$ is a reference point, which we have fixed in Case 1 \eqref{eq:ref_pt}.
Now if we consider a hypersurface $v(r)$ the induced metric in ingoing Eddington–Finkelstein coordinates is,
\begin{align}
 ds^{2}=(2v'-f(r)v'^{2})dr^{2}+r^{2}\sin^{2}\theta_{0}\Omega_{d-2}^{2}      
\end{align}
and the area is,
\begin{align}
    A_{s}=\Omega_{d-2}\sin^{d-2}{\theta_{0}}\int dr \frac{r^{d-2} \left(1-v'(r) f(r)\right)}{\sqrt{v'(r) \left(2-v'(r) f(r)\right)}}\,.
\end{align}
Again, we have a conserved quantity in terms of which we can determine $v'(r)$ given by,
\begin{align}
    v'(r)=\frac{r^{2 d} f(r)+C_{v}^2 r^4\pm\sqrt{C_{v}^2 r^{2 d+4} f(r )+C_{v}^4 r^8}}{C_{v}^2 r^4 f(r)+r^{2 d} f(r )^2}\,.
\end{align}
Along the lines of case 1, we utilize the symmetries of the geometry (owing to the fact that the diagram is infinitely extended in the spacelike direction to the left and to the right) to infer that there will be a turning point inside the black hole horizon at $t=0$ and some radial point $r=r_{*}$ that will satisfy, $v'(r)\rightarrow\infty$, this determine the conserved charge in terms of $r_{*}$,
\begin{align}
    C_{v}^{2}=-r_{*}^{2(d-2)}f(r_{*})=r_{*}^{2(d-2)}|f(r_{*})|\,.
\end{align}
As $f(r)<0$ for $0<r<r_{h}$, the conserved charge $C_{v}^{2}$ is positive. In addition to that this determines the boundary value of the spacelike extremal surface as 
\begin{align}
    v(r_{*})=\tilde{r}(r_{*})
\end{align}
and selects the $v'(r)$ solution (root) with the negative sign as the appropriate one for our purpose. Using these boundary conditions the spacelike extremal surface is found to be,
\begin{align}
    v(r)=\int_{r}^{r_{*}} dr~ v'(r)+\tilde{r}(r_{*})\,.
\end{align}
The turning point $r_{*}$ can be determined in terms of $r_{\text{st}}$ and boundary subregion time $T$ as,
\begin{align}
    T=\int_{r_{*}}^{r_{\text{st}}}v'(r) dr+\tilde{r}(r_{*})-\tilde{r}(r_{\text{st}})\,.
\end{align}
Notice that in this case there is a lower bound on the $r_{*}$, which is imposed from the requirement that $v'(r)$ has to be real \emph{i.e},
\begin{align}
    F(r,r_{*})\equiv f(r)r^{2(d-2)}- r_{*}^{2( d-2)} f(r_{*} )\geq0\,.
\end{align}
$F(r,r_{*})$ takes its minimum value of zero at $r=r_{*}$, so if we consider an expansion near $r=r_{*}$, then the first non-vanishing term, namely $(r-r_{*})\frac{dF(r,r_{*})}{dr}\Bigg|_{r=r_{*}}$, must be positive semidefinite. So for $r>r_{*}$, this leads to the constraint
\begin{align}
   \frac{dF(r,r_{*})}{dr}\Bigg|_{r=r_{*}}\geq0
\end{align}
which in turn leads to a lower bound on $r_{*}$. For a given set of values of dS radius $L$ and black hole mass $\mu$, we will denote this lower bound by $r_{\text{min}}$.
\begin{align}
    (d-2) \left(r_{\text{min}}^d-\mu~  r_{\text{min}}^2\right)-\frac{(d-1) r_{\text{min}}^{d+2}}{L^2}=0\,.
\end{align}
For the timelike surface connecting the two end points (situated inside the black hole horizon) of the two extremal spacelike surfaces anchored at the stretched horizon at time $T$ and $-T$ respectively. We can argue from the symmetry of SdS geometry under reflection around $t=0$ 
that the extremal surface is given by,
\begin{align}
    v(r)=\tilde{r}(r)\,.
\end{align}
Now that we have worked out the relevant extremal curves, we proceed to work out the complexity integrals,
\begin{align}
    C_{s}=2\,\Omega_{d-2}\sin^{d-2}\theta_{0}\left(\int_{r_{*}}^{r_{h}} dr~ r^{d-2} \int_{v_{t}(r)}^{v_{s}(r)} dv+\int_{r_{h}}^{r_{\text{st}}} dr~ r^{d-2} \int^{v_{s}(r)}_{\tilde{r}(r)} dv\right).
\end{align}
Again, we are compelled to proceed numerically as the integrals involved are not amenable analytically.
\begin{figure}[htb!]
    \centering
    \includegraphics[width=0.5\linewidth]{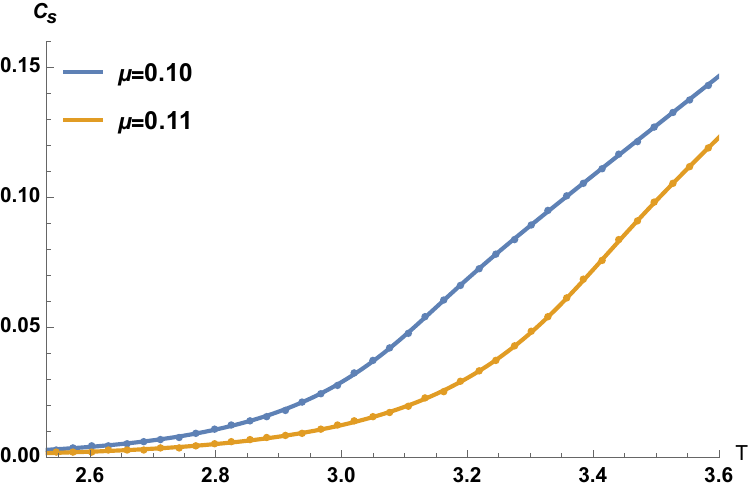}
    \caption{Plot of $C_{s}$ with boundary subregion time $T$ for $d=3$, $L=3$, $\epsilon=10^{-5}$ and different blackhole masses $\mu=0.1,0.11$ respectively (Case 2).}
    \label{fig:CvsT_BH2}
\end{figure}
\begin{figure}
    \centering
    \includegraphics[width=0.5\linewidth]{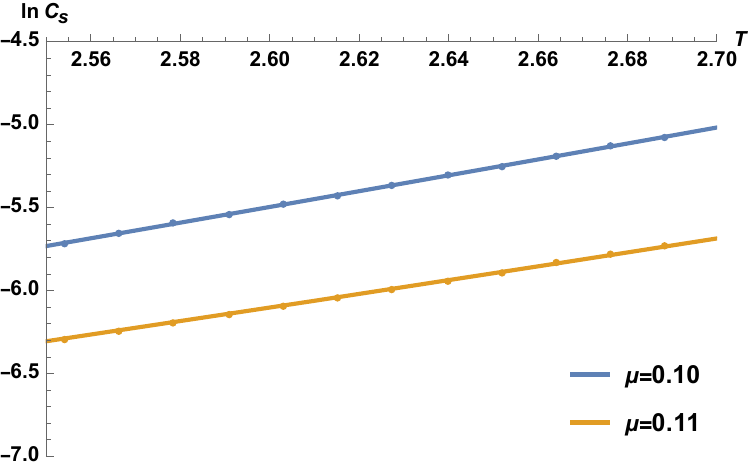}
    \caption{Early time behavior of $\ln{C_{s}}$ with boundary subregion time $T$ for $d=3$, $L=3$, $\epsilon=10^{-5}$ and different blackhole masses $\mu=0.1,0.11$ respectively. The plot shows an exponential dependence on the boundary time $T$ (Case 2).}
    \label{fig:logCvsT_earlytime_BH2}
\end{figure}
\begin{figure}
    \centering
    \includegraphics[width=0.5\linewidth]{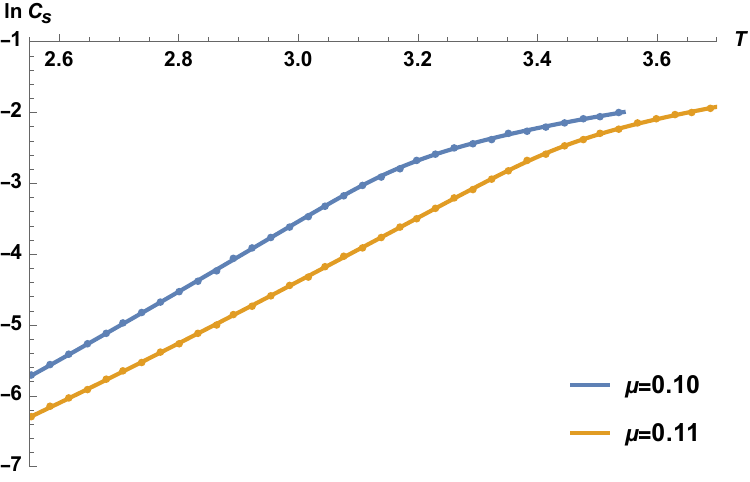}
    \caption{Plot of $\ln{C_{s}}$ with boundary subregion time $T$ for $d=3$, $L=3$, $\epsilon=10^{-5}$ and different blackhole masses $\mu=0.1,0.11$ respectively. As it is clear from the plot, it does not show hyperfast growth at a late time (Case 2).}
    \label{fig:logCvsT_BH2}
\end{figure}

We studied the behavior of timelike complexity with the subregion duration as displayed in the plots in Fig.~(\ref{fig:CvsT_BH2}). Evidently, the plots reflect a transition in the dependence of timelike complexity on the short and long boundary subregion duration. For the short boundary subregion duration, we found an exponential growth, where the characteristic growth rate decreases with the black hole mass $\mu$ as shown in Fig.~(\ref{fig:logCvsT_earlytime_BH2}). Notably, for early times, this growth rate turns out to be independent of the spacetime dimension $d$. 
For the long boundary subregion duration, we found that the complexity growth exhibits a non linear (polynomial) dependence over the boundary subregion time, as shown in Fig.~(\ref{fig:logCvsT_BH2}). Unlike the previous case, no inflationary behavior is observed. This is consistent with our expectations since the timelike complexity measures the volume of a bulk region that is extended in the black hole interior, and the black hole interior does not exhibit any inflating behavior even in dS spacetime. Furthermore, this non linear growth differs from the previously observed linear growth of spacelike volume complexity for both AdS \cite{Carmi:2017jqz} and dS black holes \cite{Aguilar-Gutierrez:2024rka}. 

\subsection{The $(2+1)$-dimensional case: Conical defect}\label{con def}
\begin{center}
 \begin{figure}
    \centering
    \includegraphics[width=0.45\linewidth]{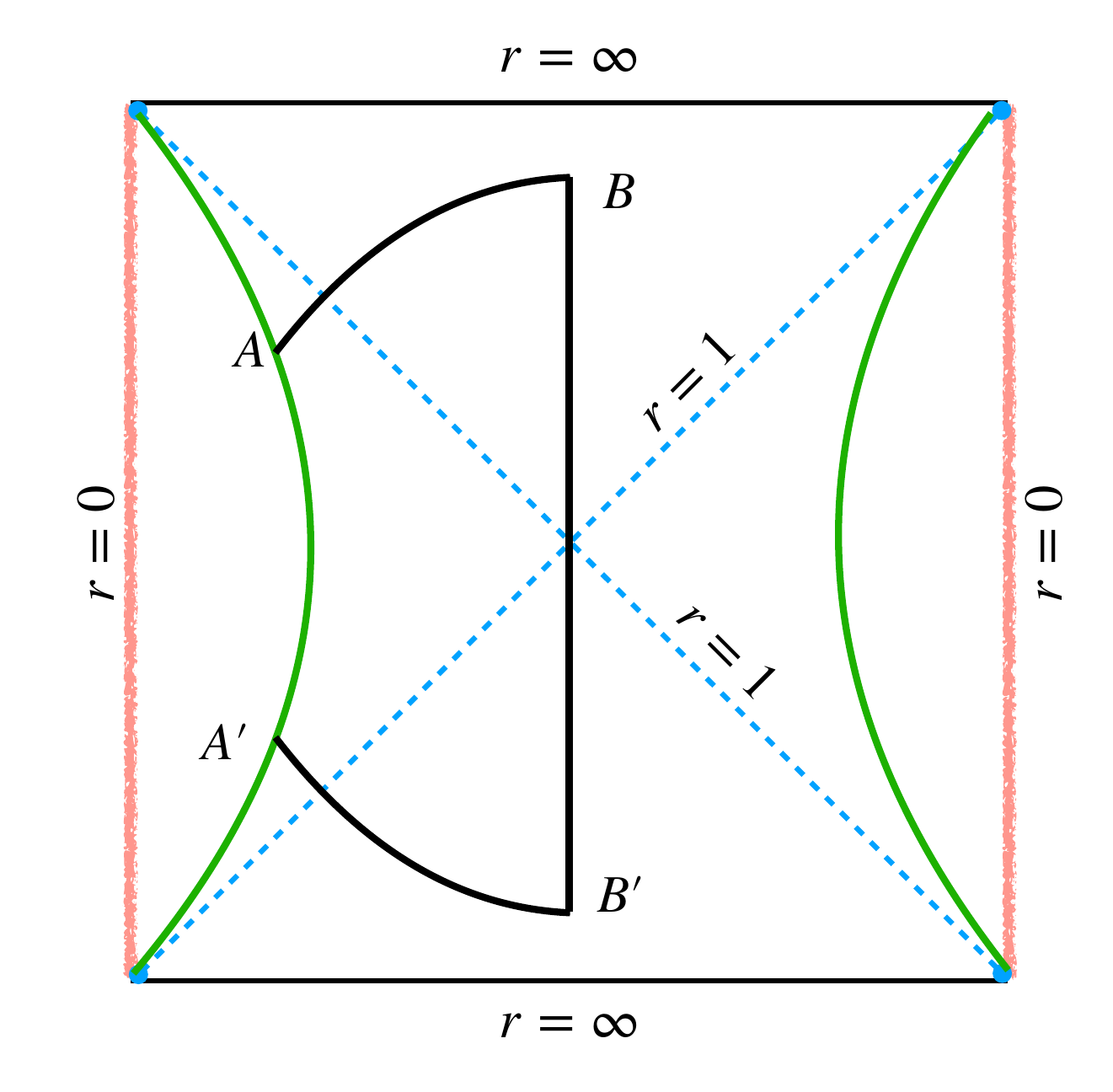}
    \caption{Penrose diagram of three dimensional de Sitter conical defect spacetime. The blue dotted lines are cosmological horizons $r=1$, while the green curves represent the streched horizons $r=1-\epsilon$. The smudged red edges at $r=0$ represent the conical defect singularity. We consider a timelike boundary subegion on the stretched horizon, the segment  $AA'$, for which the black curves $AB$, $A'B'$ represent spacelike extremal curves, while the black curve $BB'$ represent timelike extremal curve required for the computation for timelike subregion complexity.}
    \label{fig:Timelike_complexity}
\end{figure}   
\end{center}

Here we specialize to $(2+1)$-dimensions. We rewrite the metric in rescaled coordinates, $r\rightarrow l_c r, t \rightarrow t/l_c, \phi \rightarrow \varphi/l_c$, 
\begin{equation}
    ds^2=-\left(1-r^2\right) dt^2+\frac{dr^2}{1-r^2}+r^2\,d\varphi^2 \label{con_def}
\end{equation}
where it appears identical to pure dS$_3$, with the reduced range for the angular coordinate $\varphi$: $0<\varphi<2 \pi l_c$. A Penrose diagram for this conical defect spacetime is provided in Fig.~(\ref{fig:Timelike_complexity}). Thus the $r=0$ curve, demarcated in red, now represents the worldine of the conical defect singularity. It is evident that the computation of the entanglement entropy and complexity for timelike subregions, $-T<t<T, \varphi=\text{fixed}\,,$ is no different from the pure dS$_3$ case as the spacelke and timelike extremal curves remain confined to the region extraneous to the conical defect and metric is everywhere else locally pure dS$_3$. On the other hand if one were to the compute the holographic  complexity or entanglement entropy for \emph{spacelike} subregions, the reduced range of the angular coordinate will feature in the computation and the result will be different from the pure dS$_3$ case. Since this is not the agenda of our project we will skip any further discussion of that case.

\section{Discussion \& Outlook}\label{Conclusions}

Our work was motivated by the recently conjectured notion of holographic complexity for timelike subregions \cite{Alishahiha:2025xml} as a novel measure of entanglement across subregions which are separated in time, and also by the frequent appearance of timelike oriented extremal (hyper)surfaces in de Sitter spacetime as holographic duals of entanglement in the dS/CFT correspondence \cite{Doi:2022iyj,Doi:2023zaf}. We exclusively operated in the framework of static patch holography (instead of the dS/CFT correspondence) where the holographic dual field theory is supported on a timelike boundary (stretched horizon) and bears greater resemblance to the well understood AdS/CFT duality setting. In addition to providing insights about the newly proposed notion of timelike subregion (volume) complexity, our study also provides insights about the boundary dual field theory. Our aim was to work out the timelike subregion complexity for pure dS and the (infinitely extended) SdS black hole geometries following the prescription \cite{Alishahiha:2025xml}. Before conducting these studies, in order to facilitate a juxtaposition  with conventional (spacelike) volume complexity, we first revisited the analysis of \cite{Jorstad:2022mls} via numerical means. Apart from reproducing the hyperfast growth phenomenon noted in earlier studies, whereby the full volume complexity blows up in at some finite time (the critical time, $T_{\textrm{max}}$) in a power law divergence, we uncovered a previously unnoticed feature, namely that of an exponential growth of the full volume complexity at times much shorter in comparison to the critical time $\mathcal{C}(T)\sim e^{T},\,T \ll T_{\textrm{max}}$. As a representative case, when the stretched horizon cutoff is set at $\epsilon =10^{-7}$, the critical time turns out to be $T_{\textrm{max}}\approx 8.4$, and the exponential growth of holographic volume complexity persists till $T\approx 6.5$ (all length and time scales in units of the dS radius). A striking aspect of this exponential growth phase is that the growth rate is independent of the spacetime dimensionality. Contrast this with the power-law divergence of for the hyperfast growth regime, where the exponent does depend on the spacetime dimensionality, $\mathcal{C}(T)\sim \left(T_{\textrm{max}}-T\right)^{-(d-1)}$. An exponential growth is known to take place for Krylov complexity as has been recorded in the literature, e.g. \cite{Parker:2018yvk,2023JHEP...08..099B, Anegawa:2024yia,Dymarsky:2021bjq,Avdoshkin:2022xuw,Camargo:2022rnt}. Quantum computational complexity, on the other hand, is expected to grow linearly with time \cite{Susskind:2014moa}. This is perhaps evidence that in the static patch holography dictionary, the complexity in the ``Complexity $=$ Volume" duality is Krylov complexity instead of quantum computational (Nielsen) complexity. See \cite{Rabinovici:2023yex, Aguilar-Gutierrez:2024nau, Ambrosini:2024sre,Heller:2024ldz,Aguilar-Gutierrez:2026ogo} regarding the holographic dual (length) for Krylov complexity in the context of the AdS$_2$ JT-gravity-DSSYK duality. In our previous study of complexity for conformal fields in de Sitter space \cite{Parihar:2026rce} we also encountered exponential growth of complexity, and the underlying reason was obvious: the field theory was placed on a time-dependent exponentially inflating background, implying an exponential growth in the number of degrees of freedom, thereby accounting for the exponential growth in complexity (a measure of entanglement). However, here the field theory is placed on the stretched horizon, a static background, and so the exponential growth in complexity implies an exponential growth in entanglement either due to some inherent exponential scaling of interaction strength or nonlocal nature of the interactions. Incidentally, the hyperfast growth phenomenon \cite{Susskind:2021esx}, manifested in the bulk by the inflationary expansion of comoving volumes outside the static patch, has been conjectured to be due to the dual boundary field theory degrees of freedom displaying ``hyperfast scrambling" behavior \emph{also} due to extremely non-local interactions (all-to-all interactions). However, this is not the only possibility. A more conservative point of view is that there is nothing exotic or nonstandard about the field theory; it is based on local k-body interactions, but the hyperfast growth is an artifact of the holographic probe itself, i.e., volume complexity. There is an enormous degree of ambiguity in defining the bulk holographic dual of complexity \cite{Belin:2021bga, Belin:2022xmt}, and when one considers alternative bulk geometric constructions as duals for complexity, e.g., in \cite{Aguilar-Gutierrez:2023zqm,  Aguilar-Gutierrez:2023tic,Aguilar-Gutierrez:2023pnn, Mohan:2025aiw}, the hyperfast growth disappears, replaced by exponential growth or linear growth. Since all these phenomena concern the evolution of entanglement of the dual field theory, we were motivated to consider the timelike subregion complexity proposal of \cite{Alishahiha:2015rta} for the dS static patch holography, as it is tailor-made to capture timelike entanglement. If these features of early-time exponential growth and late-time hyperfast growth are universal, timelike subregion complexity must bear their signatures as well. With this aim in mind, we went on to work out the holographic timelike subregion complexity for dS. We first considered the case of pure dS${_3}$ spacetime, where analytic results were obtained, and then generalized our treatment to pure dS spacetimes of higher spacetime dimensions via a numerical treatment. The numerics were carried out in two different coordinate systems (Kruskal and Eddington-Finkelstein) so as to establish the robustness of the numerical estimates. The timelike subregion volume complexity is a function of the extension in the timelike direction, i.e., the duration, $T\geq t \geq-T$, namely $\mathcal{C}=\mathcal{C}(T)$. We found that the timelike subregion volume complexity shares the same traits as full volume complexity, namely exponential growth regime for shorter durations followed by a hyperfast growth regime for long durations with the complexity blowing up at a finite subregion size (duration) which is \emph{identical} to the the maximal critical time $T_{\textrm{max}}$ whereby full volume complexity also diverges. This can perhaps be explained by the fact that in the timelike subregion complexity prescription, one has to first work out the spacelike extremal surfaces anchored at the endpoints of the timelike extended subregion of the boundary (stretched horizon), much akin to full volume complexity where the maximal volume hypersurface is also spacelike anchored at a boundary time (albeit the surface is codimension 2 in the first case and codimension 1 in the second case).  For subregion durations shorter than $T_{\textrm{max}}$, the timelike subregion volume complexity grows exponentially with a growth-constant independent of the spacetime dimensionality, $\mathcal{C}(T)\sim e^{2T}$. Note that the growth constant is twice that of full volume complexity. In this regard, we would like to point out that the growth rate tracks the codimensionality of the hypersurface on the boundary where the bulk extremal spacelike curve is anchored. For full volume complexity, the bulk extremal curve is an $d$-dimensional hypersurface that intersects the stretched horizon on a $(d-1)$ dimensional hypersurface, i.e., a codimension $1$ hypersurface, and the growth constant is $1$. On the other hand, for the timelike subregion complexity case, the spacelike bulk extremal hypersurface is $(d-1)$-dimensional, which intersects the stretched horizon on a $(d-2)$-dimensional hypersurface, i.e. a codimension $2$ hypersurface and the growth constant is $2$ as well. For long subregion durations, close to the maximal subregion duration, we found that the timelike subregion volume complexity exhibits a hyperfast growth, $\mathcal{C}(T)\sim \left(T_{\textrm{max}}-T\right)^{-(d-2)}$. The exponent of this power law divergence is one less than that of full volume complexity. One can perhaps account for this from the fact that the number of boundary spacelike dimensions at the extremities of the timelike oriented subregion where the bulk extremal spacelike surface is anchored is $(d-2)$, while for full volume complexity the number of boundary spacelike dimensions where the bulk extremal spacelike surface is anchored is $(d-1)$, i.e., the hyperfast growth or divergences is contributed by the boundary spatial extensions. Thus, overall, we find that timelike subregion volume complexity captures the same qualitative features as that of full volume complexity with few minor distinctions. In this respect, our work could be regarded as corroborating evidence strengthening the case for the validity or the viability of the timelike subregion volume complexity idea.
\\

Next, we expanded the scope of our investigation from pure dS to an asymptotically dS background, namely the extended Schwarzschild de Sitter black holes. Several conjectured holographic duals of complexity were investigated for the Schwarzschild de Sitter (SdS) black hole background in the setup of static patch holography in the literature \cite{Aguilar-Gutierrez:2024rka}, where various different configurations of stretched horizons (boundary) were considered. For the two specific configurations of stretched horizons, namely when the stretched horizon lies very near to the cosmological horizon, it is referred to as Case 1 (Section~(\ref{subsec: case1})), and the case when the stretched horizon lies very near to the black hole horizon is referred to as Case 2 (Section~(\ref{subsec: case2})). The study \cite{Aguilar-Gutierrez:2024rka} found quite distinct late-time behaviors of the volume complexity growth for the two cases. In particular, in Case 1, the volume complexity exhibits a hyperfast growth, thereby capturing the inflating behavior of de Sitter spacetime. No dependence on the black hole mass parameter $\mu$ was found. In contrast, in Case 2, the volume complexity displays the conventional linear growth at late times that is observed for AdS black holes, corresponding to the linear growth of the Einstein-Rosen bridge (ERB) volume.  Our study of timelike subregion complexity for the long boundary durations in Case 1 is in agreement with the behavior of full volume complexity, i.e., the timelike subregion volume complexity for Case 1 also exhibits a hyperfast growth with an exponent dependent on both the mass parameter of the black hole and spacetime dimensionality. The dependence of spacetime dimension is the same as was observed for the case of pure de Sitter spacetime \emph{i.e} $(d-2)$. While the explicit functional dependence on the mass parameter of black hole could not be determined due to the numerical nature of our analysis, it is evident from the numerical results that the exponent is a monotonically decreasing function of the black hole mass $\mu$ when $T$ reaches $T_{\text{max}}$ even though this dependence on mass parameter disappears when the gap between $T$ and $T_{\text{max}}$ increases and of order $\mathcal{O}(10^{-3})$ for $\epsilon=10^{-3}$. The maximal time for the hyperfast growth also depend on the black hole mass parameter $\mu$ (in addition to the usual dependence on the stretched horizon cut off $\epsilon$) as we have established numerically, and in particular is a monotonically increasing function of $\mu$. In other words, black hole mass delays the hyperfast growth regime. In addition to this late-time behavior, we have also studied the dependence of complexity on boundary subregion time in short boundary durations. This study exhibits an exponential behavior with the exponent depending on the mass parameter of the black hole. Again, due to the numerical nature of our analysis in this situation, exact functional dependence on the mass parameter is not exactly determined, but it is evident that the exponent is a monotonically decreasing function of the mass parameter $\mu$. As a consistency check, we also reproduce its pure de Sitter value of two when $\mu$ vanishes.  So, for Case 1, the results of our investigation were qualitatively similar to those for the pure dS case. But for Case 2, our results were in marked contrast to the pure dS case. As mentioned before, the origin of this contrast in the qualitative behavior owes to the fact that in Case 2, the stretched horizon lies near the black hole horizon, and the extremal surfaces extend into the black hole interior. We also found  that the behavior of timelike subregion complexity for the large boundary durations is rather different from the previous observations for the full volume complexity. The timelike subregion volume complexity exhibits nonlinear growth (mix of powers) rather than purely linear growth. We believe this nonlinear power law growth phenomenon indicates that the timelike volume subregion complexity does not correspond to the ERB volume. Intuitively, it makes sense since the timelike subregion volume complexity not only involves an ERB-like spacelike hypersurface, but a timelike hypersurface as well, and in the black hole interior, such timelike segments grow nonlinearly. We have also studied the dependence of complexity on boundary subregion time in short boundary subregion durations. This study exhibits an exponential behavior with the exponent depending on the mass of the black hole, and is a monotonically decreasing function of the mass parameter $\mu$. However, the growth constant for the exponential growth turned out to be independent of spacetime dimensionality, just like in the pure dS situation.
\\

Finally, there remain several open questions and interesting avenues of further investigations; we discuss a few of them here. The first one is regarding the universality of the full volume complexity, in particular, its early-time exponential growth phenomenon. As discussed earlier in this section, the late-time hyperfast growth gets replaced when one considers alternative prescriptions for complexity, i.e. is not a universal feature in all prescriptions of complexity. Does the exponential growth phenomenon at subcritical times suffer the same fate? It would be interesting to check what happens to the early time exponential regime in some of the alternative holographic complexity proposals, especially for those where the hyperfast growth disappears. Another related issue is the generalization of the timelike subregion complexity to other holographic prescriptions for complexity, e.g., for action complexity \cite{r14,r15} or for CV 2.0 \cite{Couch:2016exn} or for even more general complexity \cite{Belin:2021bga, Belin:2022xmt}. This is an issue that is quite independent of the context of dS holography, i.e., it will be relevant to holography in general backgrounds such as AdS or flat. For dS static patch holography, an interesting scenario which can be investigated is the timelike subregion volume complexity for the lower-dimensional dS$_2$ case in the context of JT-gravity, where one also has a dilaton background turned on.  In such a case, one must consider a modified volume complexity functional along the lines of those considered for the holographic volume complexity in holographic backgrounds with a running dilaton, see e.g. \cite{Brown:2018bms, Chakraborty:2020fpt,Katoch:2022hdf, Bhattacharya:2023drv,Balusu:2025kht}. A related issue is regarding the volume complexity and holographic measures of timelike entanglement for theories which are holographic duals of asymptotically flat spacetimes, particularly in the setup of Flat/Carroll CFT duality \cite{Ruzziconi:2026bix} \footnote{Circuit complexity for Carrollian conformal field theories is investigated in \cite{Bhattacharyya:2023sjr}.}. This case is particularly interesting because the boundary is a null manifold (future null infinity). For an interesting recent attempt see \cite{Jorstad:2026jlg}. Actually, this is an even more mysterious issue in the Celestial holography framework since the boundary is purely Euclidean, namely the celestial sphere, and thus the problem of timelike entanglement is not naturally defined or well-posed. Another interesting direction to pursue would be to look at timelike subregion complexity for nonlocal field theories with a holographic (super)gravity dual, e.g Little string theories dual to asymptotically flat linear dilaton backgrounds \cite{ Chakraborty:2020fpt,Katoch:2022hdf}, or warped CFTs \cite{Bhattacharyya:2022ren}, or SYM with dipole deformations or other noncommutative deformations \cite{Balusu:2025kht}. Finally one would like to expand the scope of our studies on the timelike subregion volume complexity of SdS black holes to incorporate effects of rotation and charges. It would be particularly interesting to see, for the situation where the boundary (stretched horizon) is placed near the black hole horizon, whether the power law growth of timelike subregion volume complexity we obtained for the SdS case persists in the presence of angular momentum and charge or gets modified even further.

\acknowledgments
The work of AB, SP, and SR is supported by the Core Research Grant (CRG) of the Anusandhan National Research
Foundation (ANRF), Department of Science and Technology (DST), CRG/2023/001120 (``\emph{Many facets
of complexity: From chaos to thermalization}"). The work of MKH and SR is partially supported by the IIT Hyderabad research funds RDF/IITH/F171/SR. A.B. also acknowledges the associateship program of the Indian Academy of Sciences, Bengaluru, and support from the Indian Institute of Technology Gandhinagar and a generous donor through the Singheswari and Ram Krishna Jha Chair. SR thanks the organizers of \href{https://sites.google.com/view/holo-asia2026/home}{``Holo-Asia 2026" workshop on quantum information, 
matter and gravity} (June 25-29, 2026) in Jeju, South Korea, where a preliminary version of this work was presented. 


\appendix

\section{Check on numerical analysis for the pure dS case in Eddington-Finkelstein coordinates} \label{A1}
In Eddington-Finkelstein coordinates, the area functional to be extremised is given by,
\begin{align}
    A=\Omega_{d-2}\int dr ~r^{d-2} \sqrt{-u'(r) \left(f(r) u'(r)+2\right)}\,.
\end{align}
For three-dimensional de Sitter, we can perform the extremization analytically. After the extremization we found the spacelike extremal surface as,
\begin{align}
    u_{s}(r)=-2 \tanh ^{-1}\left(\frac{r \left(r_{*}-\sqrt{r_{*}^2-1}\right)}{r_{*}+\sqrt{r_{*}^2-r^2}}\right)
\end{align}
where $r_{*}$ is determined by the boundary condition on the stretched horizon,
\begin{align}
    r_{*}=\frac{1-\epsilon }{\sqrt{\epsilon ^2 \cosh ^2(T)-2 \epsilon  \cosh ^2(T)+1}}\,.
\end{align}
Therefore, the area of the spacelike extremal surface is,
\begin{align}
    A_{s}=\tan ^{-1}\left(\frac{\sqrt{r_{*}^2-(1-\epsilon )^2}}{1-\epsilon }\right)\,.
\end{align}
The timelike extremal surface is found to be,
\begin{align}
    u_{t}(r)=-\tanh ^{-1}\left(\frac{1}{r}\right)\,.
\end{align}
and the area of the timelike extremal surface is,
\begin{align}
    A_{t}=\tanh ^{-1}\left(\frac{\sqrt{r_{*}^2-1}}{r_{*}}\right)\,.
\end{align}
The timelike complexity is found to be,
\begin{align}
    C_{s}=2\Omega_{d-2}\int_{1}^{r_{*}}dr~(u_{s}(r)-u_{t}(r))+2\Omega_{d-2}\int_{1-\epsilon}^{r_{*}}dr (u_{s}(r)+\tanh ^{-1}(r))\,.
\end{align}
The resultant answer is in exact agreement with the results found for timelike complexity in Kruskal coordinates as shown in Table~(\ref{tab:EFandKruskal}). Although in $d>2$ we have to rely on numerical methods, the numerical results obtained in EF coordinates are found to be in excellent agreement with those computed in Kruskal coordinates for $d=3,4,5$. This consistency between two independent coordinate systems further strengthens confidence in the validity and robustness of the results.
\begin{table}[]
    \centering
    \begin{tabular}{|c|c|c|}
    \hline
      T& $C_{s}^{\text{EF}}$&$C_{s}^{\text{Kruskal}}$\\
\hline
1& 2.81$\times10^{-7}$ & 2.81$\times10^{-7}$ \\
 \hline
2& 1.56$\times10^{-6}$  & 1.56$\times10^{-6}$  \\
 \hline
3& 1.04$\times10^{-5}$  & 1.04$\times10^{-5}$  \\
 \hline
4& 7.5$\times10^{-5}$  & 7.5$\times10^{-5}$  \\
 \hline
5& 5.5$\times10^{-4}$  & 5.5$\times10^{-4}$  \\
 \hline
 6&4.08$\times10^{-3}$  & 4.08$\times10^{-3}$  \\
 \hline
    \end{tabular}
    \caption{A comparison between timelike complexity obtained in EF coordinates and Kruskal coordinates for $dS_{3}$ for $\epsilon=10^{-7}$. }
    \label{tab:EFandKruskal}
\end{table}
\section{A late time analysis of the timelike subregion volume complexity for SdS Black holes for Case 2} \label{A2}
In the late time limit, in addition to the numeric analysis, we can study the behavior of complexity growth analytically. To proceed with that, we first consider the boundary subregion time,
\begin{align}
        T=&\int_{r_{*}}^{r_{\text{st}}}v'(r) dr+\tilde{r}(r_{*})-\tilde{r}(r_{\text{st}})\,,\nonumber\\
        =&\int_{r_{*}}^{r_{\text{st}}}dr\left(\frac{1}{f(r)}-\frac{\sqrt{ r_{*}^{2 (d-2)} |f(r_{*} )|}}{f(r)\sqrt{(r_{*}^{2 (d-2)} |f(r_{*} )|+r^{2( d-2)} f(r ))}}\right)-\int_{0}^{r_{*}}\frac{dr}{|f(r)|}-\int_{r_{0}}^{r_{st}}\frac{dr}{|f(r)|}\,,\nonumber\\
         =&\int_{r_{*}}^{r_{\text{st}}}dr\left(\frac{1}{f(r)}-\frac{\sqrt{ r_{*}^{2 (d-2)} |f(r_{*} )|}}{f(r)\sqrt{F(r,r_{*})}}\right)-\int_{0}^{r_{*}}\frac{dr}{|f(r)|}-\int_{r_{0}}^{r_{st}}\frac{dr}{|f(r)|}\,.
\end{align}
We consider the limiting case of $r\rightarrow r_{*}$, then,
\begin{align}
    T\sim&\int_{r_{*}}dr\left(-\frac{\sqrt{ r_{*}^{2 (d-2)} |f(r_{*} )|}}{f(r)\sqrt{F(r,r_{*})}}\right)\,,\nonumber\\
    \sim&\int_{r_{*}}dr\frac{r_{*}^{d-2}}{\sqrt{\frac{dF}{dr}\Big|_{r=r_{*}}(r-r_{*})+\frac{1}{2}\frac{d^{2}F}{dr^{2}}\Big|_{r=r_{*}}(r-r_{*})^{2}+...}}\,.
\end{align}
Now if we consider $r_{*}=r_{\text{min}}$ then we get,
\begin{align}
    T\sim -\frac{r_{\text{min}}^{d-2}\,\ln\delta}{\sqrt{\frac{1}{2}\frac{d^{2}F}{dr^{2}}}\Big|_{r=r_{\text{min}}}}\,.
\end{align}
Now we consider the timelike complexity in the same limiting case,
\begin{align}
    C_{s}\sim&\int_{r_{*}} dr~ r^{d-2} (v_{s}(r)-\tilde{r}(r))\,,\nonumber\\
    C_{s}\sim&\int_{r_{*}} dr~ r_{*}^{d-2} \int_{r_{*}}^{r}dy \frac{r_{*}^{d-2}}{\sqrt{\frac{dF}{dy}\Big|_{y=r_{*}}(y-r_{*})+\frac{1}{2}\frac{d^{2}F}{dy^{2}}\Big|_{y=r_{*}}(y-r_{*})^{2}+...}}\,.
\end{align}
Considering $r_{*}=r_{\text{min}}$ then we get,
\begin{align}
    C_{s}\sim& \int_{r_{\text{min}}+\delta}dr~\frac{r_{\text{min}}^{2(d-2)}}{\sqrt{\frac{1}{2}\frac{d^{2}F}{dr^{2}}}\Big|_{r=r_{\text{min}}}}\Big(\ln(r-r_{\text{min}})-\ln \delta\Big)\,,\nonumber\\
    C_{s}\sim&-\frac{r_{\text{min}}^{2(d-2)}}{\sqrt{\frac{1}{2}\frac{d^{2}F}{dr^{2}}}\Big|_{r=r_{\text{min}}}}\Big(\delta\ln\delta-\delta-(r_{\text{min}}+\delta)\ln\delta\Big)\,,\nonumber\\
    C_{s}\sim&\frac{r_{\text{min}}^{2(d-2)}}{\sqrt{\frac{1}{2}\frac{d^{2}F}{dr^{2}}}\Big|_{r=r_{\text{min}}}}\Big(\delta+r_{\text{min}}\,\ln\delta\Big)\,.
\end{align}
In $\delta\rightarrow0$ limit,
\begin{align}
    C_{s}\sim& \frac{r_{\text{min}}^{2(d-2)}\, r_{\text{min}}\,\ln\delta}{\sqrt{\frac{1}{2}\frac{d^{2}F}{dr^{2}}}\Big|_{r=r_{\text{min}}}}\,,\nonumber\\
    C_{s}\sim& r_{\text{min}}^{d-1}\,T\,.
\end{align}
The presence of $r_{\text{min}}^{d-1}$ in the above expression clearly indicates that even in the long boundary subregion duration (late time) the timelike complexity does not show linear growth with time.

\bibliographystyle{JHEP}
\bibliography{references.bib}
\end{document}